\documentclass[fleqn,usenatbib]{mnras}
\usepackage{newtxtext,newtxmath}
\usepackage[T1]{fontenc}

\DeclareRobustCommand{\VAN}[3]{#2}
\let\VANthebibliography\thebibliography
\def\thebibliography{\DeclareRobustCommand{\VAN}[3]{##3}\VANthebibliography}

\usepackage{graphicx}	
\usepackage{amsmath}	
\usepackage{epsfig}
\usepackage{epstopdf}
\usepackage{float}
\usepackage{subfigure}
\usepackage{hyperref}
\usepackage{multirow}
\usepackage{times}  
\usepackage{soul}
\usepackage{float}
\usepackage{comment}
\usepackage{hyperref}
\usepackage[title]{appendix}

\def\HI{{H~{\sc i} }}
\def\HII{{H~{\sc ii} }}
\def\Msun{\rm  M_{\sun} \ }
\def\kms{\ \rm km \ s^{-1}}
\def\12CO{^{12}\rm CO}

\title[Vertically constrained H~{\small I} in W41--W44 region]{Observational evidence for local vertical constraining of \HI by molecular cloud complexes}

\author[Nandakumar et.al ]
{Meera Nandakumar$^{1}$\thanks{Email: meeran@iisc.ac.in}, 
Nirupam Roy$^{1}$, \thanks{Email:nroy@iisc.ac.in},  
Chanda J. Jog$^{1}$ \thanks{Email:cjjog@iisc.ac.in}, 
Karl M. Menten$^{2}$\thanks{Email:kmenten@mpifr-bonn.mpg.de}
\\$^{1}$ Department of Physics, Indian Institute  of Science, Bengaluru, 560012  India. 
 \\$^{2}$ Max-Planck-Institut f\"{u}r Radioastronomie, Auf dem H\"{u}gel 69, 53121 Bonn, Germany
}

\date{Accepted XXX. Received YYY; in original form ZZZ}

\pubyear{2023}

\begin{document}
\label{firstpage}
\pagerange{\pageref{firstpage}--\pageref{lastpage}}
\maketitle

\begin{abstract}
A massive molecular cloud complex represents local gravitational potential that can constrain the vertical distribution of surrounding stars and gas. This pinching effect results in the local corrugation of the scale height of stars and gas which is in addition to the global corrugation of the mid-plane of the disc. For the first time, we report observational evidence for this pinching on the \HI vertical structures in the Galactic region ($20^{\circ}< l<40^{\circ}$), also called W41--W44 region. The \HI vertical distribution is modelled by a double Gaussian profile that physically represents a narrow dense gas distribution confined to the mid-plane embedded in a wider diffuse \HI. We find that the estimate of the \HI scale height distribution of wider components, shows corrugated structures at the locations of molecular complexes,  as theoretically predicted in literature. While the narrow component is less affected by the pinching, we found a hint of the disc being disrupted by the active dynamics in the local environment of the complex, e.g., supernova explosions. Molecular complexes of mass of several $10^6 \Msun$, associated with the mini-starburst region W43 and the supernova remnant W41 show the strongest evidence for the pinching; here a broad trough, with an average width of $\sim400$ pc and height $\sim300$ pc, in the disc thickness of the wider component is prominently visible. Searching for similar effect on the stars as well as in the location of other complexes in the Milky Way and other galaxies will be useful to establish this phenomenon more firmly.  
\end{abstract}


\section{Introduction}
\label{Intro}
The vertical distribution of stars and gas in disc galaxies is a well-explored topic due to its significance in our  understanding of the galactic potential and dynamical effects. Evidence from an enormous amount of Galactic and extra-galactic observations established that the vertical distribution of stars and gas rather being smooth and uniform, in fact, is generally uneven in a complex manner, which is caused by multiple phenomena like disc flaring, warping, disc bending and others \citep{1957AJ.....62...90B,1969ApJ...155..747H,1974Ap&SS..27..323Q,1990ARA&A..28..215D,2020MNRAS.495.3705N,2022MNRAS.513.3065N}. Though most of these are identified as global phenomena that lead to perturbations of  the entire disc structure, a much smaller number of studies has focused on the dynamical effects that perturb the stellar and gas vertical distribution locally, meaning over a certain region of the disc. 
\cite{ChandaP1} investigate one such local dynamical effect that massive molecular cloud complexes are creating on the vertical distribution of stars and gas, which they refer to as the constraining or ``pinching'' of the disc.
Much of the Milky Way's interstellar molecular gas is contained in Giant Molecular Clouds (GMCs) that have masses of $10^4$ to $10^6~\Msun$ and  sizes from tens to hundred parsecs  
\citep{1986ApJ...305..892D,1987ASSL..134...21S}. These are further segregated into large molecular cloud complexes or clusters \citep{1986Rivolo}.
\cite{ChandaP1} theoretically modelled the self-consistent response of the stellar and gas disc to the gravitational potential of such a complex and showed that the scale height of both the stars and \HI gas  becomes reduced by a factor of $\sim 3$ compared to the case in which the complex was absent. They show that the constraining effect in the gas can extend over a distance of $\sim 500$ pc from the complex centre along the mid-plane. 
Given that the magnitude of the resulting broad trough-like features  
in the scale height distribution is significant, it should make these features accessible to observations.

In this work, we investigate this constraining effect  
 on the vertical distribution of \HI gas surrounding molecular clouds in various
 complexes in the Milky Way.
For this we first briefly introduce the vertical distribution in a typical galactic disc.
Knowing the average vertical distribution of various disc components is crucial in understanding  galactic dynamics, however, obtaining the three-dimensional distribution of the interstellar medium (ISM) through observations 
 has its  challenges in both the Milky Way and in external galaxies. Highly inclined edge-on galaxies are best for probing the vertical structures of the disc by photometry. Modelling the vertical distribution by an exponential function, \cite{2014MNRAS.441..869D} reported an average scale height of the stellar disc of $\sim 510$ pc for 12 edge-on galaxies. Using another sample of 34 edge-on spiral galaxies from \cite{2002MNRAS.334..646K}, they obtained similar estimates of the stellar scale height of $570$ pc. Inspecting the surface brightness profile determined in different optical bands reveals growing evidence for an increase in the stellar scale height along their major axes \citep{1997A&A...320L..21D,2019RNAAS...3...74T,2020MNRAS.493.5464K}. A flaring stellar disc is shown to be a generic feature from theoretical modelling of a multi-component galactic disc in hydrostatic balance \citep{2002A&A...390L..35N,2002A&A...394...89N,2018A&A...617A.142S,2019A&A...628A..58S}.
 \cite{2014MNRAS.441..869D} estimate an average scale height of $ 460$ pc for dust emission in the same sample of $12$ edge-on galaxies, similar to the values found in previous studies \citep{ 1999A&A...344..868X, 2007A&A...471..765B}.
 \cite{2022RAA....22h5004Z} studied the vertical distribution of the \HI gas disc of  15 highly inclined nearby galaxies from the Continuum Halos in Nearby Galaxies -
an EVLA Survey \citep[CHANG-ES;][]{2012AJ....144...43I}.
The photometric \HI scale height they measured for this sample ranges from $\sim 700$ pc to $2$ kpc.
 Assuming the disc to be in hydrostatic equilibrium, the \HI scale height of low inclination galaxies can also be estimated from observables like gas velocity dispersion and surface density \citep{1995AJ....110..591O, 1996AJ....112..457O,2019A&A...622A..64B}.
 The \HI scale height of several spiral and dwarf galaxies has been estimated by
 \cite{2020MNRAS.495.2867P}  and \cite{2020MNRAS.499.2063P}, who found values around a few 100 parsecs near the centre, which extended to a few kiloparsecs in  outer regions. A recent study 
 by \cite{2021ApJ...916...26R} on \HI rich galaxies selected from the BLUEDISK, THINGS, and VIVA surveys found that the average scale height of \HI within the extent of the stellar disk is $\sim 400$ pc. A similar analysis has been carried out to estimate the molecular scale height by a few authors \citep{2019ApJ...882....5W,2019MNRAS.484...81P,2019A&A...622A..64B}.
 Molecular gas is found to be much more constrained in thin discs around the mid plane, with the H$_2$ distributions having scale heights ranging from
 $\sim 20$ pc to $100$ pc in the inner parts that flare up to $\sim 200$ pc in the outer regions.  Physically, the scale height of a disc component is a measure of the
equilibrium between the local vertical gravitational force and the pressure
gradient, as given by the equation of hydrostatic equilibrium \citep[e.g.,][]{1977LNP....69.....R}. A realistic disc consisting of stars and gas that are gravitationally coupled within the dark matter halo has been modelled theoretically. This model gives a self-consistent vertical distribution including the scale height, for each of the disc components, in good agreement with the observed data \citep{2002A&A...394...89N,2003ApJ...588..805K,2018A&A...617A.142S}.

 In the Milky Way, the scale height of stellar and gas components is estimated directly from their number density distribution. Using the photometric data from various surveys like SDSS, XSTPS-GAC etc., numerous studies have derived the Galactic structure parameters like scale height and scale length \citep{2016ARA&A..54..529B,2017MNRAS.464.2545C}. These results suggest the presence of a thin disc and a thick disc component for the stellar distribution with exponential scale heights of $\sim250$ pc and $\sim 700$ 
 pc, respectively. 
With the advent of extensive observational data of the Galactic plane by numerous surveys at multiple wavelengths including the infrared, optical and radio ranges, multiple attempts have been made to map the three-dimensional distribution of dust and gas in the Galactic disc \citep{2001ApJ...556..181D,2003PASJ...55..191N,2006A&A...459..113M,2006ApJ...643..881L,2006A&A...453..635M,2011AJ....142...44J,2017A&A...607A.106M,2018ApJ...858...75L,2022MNRAS.516..907S}.
The vertical structure of the Galactic \HI disc, studied by various authors, is found to have a thickness that remains smooth in an inner radius of $< 5$ kpc and increases radially towards the outer Galaxy \citep{1962dmim.conf....3O,1986ApJ...301..380L,1990ApJ...365..544B,1990ARA&A..28..215D,1995ApJ...448..138M,  2002A&A...394...89N,2003PASJ...55..191N, 2008A&A...487..951K,2009ARA&A..47...27K}. Additionally, the vertical distribution of gas is observed to have complex components like an extended halo. \cite{1990ARA&A..28..215D}    
approximate the \HI vertical distribution within Galactocentric radii, 
$R$, of $0.4 R_{\sun} \lesssim R \lesssim R_{\sun}$, by a combination of a double Gaussian  with best fitting Full Width at Half Maximum (FWHM) of $ 212$ pc, $530$ pc respectively and  an exponential function with scale height of $403$ pc. $R_{\sun}$ is the distance of the Sun from the Galactic center. 
Like in external galaxies, the Galactic molecular hydrogen has a much thinner 
distribution with average FWHM values starting from $\sim 60$ pc near the Galactic centre that increase radially up to $140$ pc beyond a radius of $2$ kpc \citep{ 1984ApJ...276..182S,1987ApJ...322..706D,1988ApJ...327..139C,1988ApJ...324..248B,1994ApJ...433..687M,2001ApJ...547..792D,2006ApJS..163..145J,2015ARA&A..53..583H}. 
From the three-dimensional analysis of Milky Way dust, \cite{2018ApJ...858...75L} find the global scale height of the dust emission to be $103$ pc, which agrees with previous results \citep{2001ApJ...556..181D,2006A&A...453..635M,2011AJ....142...44J}.
In addition to the scale height of the neutral medium, using multiple interstellar tracers, the vertical distributions of the warm ionized medium and that of the hot ionized medium in the Galactic disc were found to have exponential scale heights
of $\sim 300  - 1000 $ pc and  $\sim 2$ kpc - $5$ kpc, respectively  
\citep{1984ApJ...282..191R,1992ApJS...83..147S, 1994ApJ...434..599S,1997AJ....113.2158S, 2000ApJ...538L..27S,2001RvMP...73.1031F, 2014A&A...564A.101L}.

Most of the above-mentioned papers 
estimate a global value or an azimuthally averaged radial profile of the scale height. However, to detect a localized phenomenon like pinching,
an estimation of the disc thickness at a specific location or region of the Galaxy is essential.
Despite the availability of numerous high-sensitivity Galactic gas surveys and having the advantage of resolving the structures much  better than in external galaxies, 
three-dimensional mapping of the gas density distribution in the Milky Way is challenging due to the location of the Sun.
The geometrical transformations from the observed brightness temperature, which is a function of Galactic latitude, longitude and radial velocity to a coordinate system that describes the Milky Way disc (e.g., ($R, \phi, z $)), strongly 
depends on the shape of the Milky Way rotation curve \citep{2003PASJ...55..191N,2007A&A...469..511K,2008A&A...487..951K}. However, this transformation requires an accurate determination of the Milky Way's rotation curve  \citep{2007A&A...469..511K}.    
Even with a better estimate of the rotation curve \citep{Reid2019}, due to geometrical reasons, the distance along the line of sight direction frequently cannot be uniquely determined in the inner Galaxy due to the near-far distance ambiguity 
 \citep[e.g.,][]{2003PASJ...55..191N}. However, at a tangent point, where the line of sight at a particular longitude is tangential to the circle of Galactocentric radius R, the distance to that point can be determined without any ambiguity. 

  In this work, we explore the vertical distribution of neutral gas in the region of Galactic longitude, $l$, from $20^{\circ}$ to $40^{\circ}$ (hereafter the W41--W44 region), which is the pilot region of the GLOSTAR \cite[A Global View of Star Formation in the Milky Way,][]{Brunthaler2021} survey and the MeGaPlug \citep[Metrewave Galactic Plane with the uGMRT,][]{Dokara2023} survey. This region hosts several well-known extended \HII regions and supernova remnants like W41, W42, W43 and W44, first identified by their radio emission \citep{Westerhout1958}.  These sources are serendipitously situated around the tangent point in the Galactic disc and hence the W41--W44 region enables us to estimate the vertical structure parameters from observations with less uncertainty. The study by \citet{Medina2019} of the ``pilot region'' of the GLOSTAR-VLA survey identifies numerous extended and compact radio sources in the $28^{\circ}<l<36^{\circ}$ range that include the star-forming complex W43 and the supernova remnant (SNR) W44.  A complementary large-scale Galactic plane survey, MeGaPluG, covering bands in the low-frequency radio regime ($< 1$ GHz) is planned for the uGMRT and in a precursor study, \citet{Dokara2023} conducted uGMRT observation of the W43/W44 region in the $300-750$ MHz frequency range. A plethora of  multi-wavelength Galactic surveys with high sensitivity have been conducted or are underway, addressing many aspects of Galactic astronomy and in particular star formation on a global scale, while detailed studies addressing 
  local dynamical effects the newly formed stars and their nascent GMCs exert on the surrounding interstellar medium (ISM) are rare.
The rest of the paper is divided as follows.  A brief review on the theoretical predictions of the pinching effect  made by \citet{ChandaP1} is given in Section~\ref{Theory}. Section~\ref{Sec:Data} presents details about the W41--W44 region and describes the  observational data used in our analysis. The method of estimating the scale height of \HI distribution in this region is explained in Section~\ref{Sec:Est}. Section~\ref{Sec:Res}
and Section~\ref{Sec:Con}
discuss the findings of our analysis.

\section{Vertical constraining effect of molecular cloud complex on stars and \HI gas}
\label{Theory}
A typical molecular cloud complex (also known as GMC complex) has a mass of $\sim 10^7 \Msun$ and a
 size of  a few 100 pc, and an oblate spheroidal shape, 
 and a central total scale height of about 120 pc \citep{1986Rivolo}. \citet{ChandaP1} noted that such a massive, extended complex dominates its local gravitational field. The average mass density of molecular gas within a complex is $\rm \sim 1 \Msun pc^{-3}$, which is six times larger than the dynamical or total mid-plane density, or the Oort limit of 0.15 $\rm \Msun pc^{-3}$\citep{1960BAN....15...45O}.
Thus 
the stars and gas in the surrounding galactic disc would respond to the additional, dominant gravitational field of the complex and would tend to get clustered towards the mid-plane, with a resulting decrease in their vertical scale heights. \citet{ChandaP1} obtained the redistributed self-consistent stellar
vertical distribution by solving together the joint Poisson equation and the hydrostatic balance equation for an isothermal stellar disc
taking account of the gravitational 
force due to stars and the complex, while keeping the surface density of the stellar disc at a given galactocentric radius constant.
The redistributed stellar distribution is shown to be constrained closer to the mid-plane, such that at the centre of the complex, the stellar mid-plane density increases by a factor of 2.6 and the corresponding vertical scale height (Half Width Half Maximum (HWHM) of the vertical stellar distribution) decreases by a factor of 3.4 compared to the undisturbed disc. Similarly the interstellar \HI
gas distribution also shows a similar constraining effect, here the \HI gas gravity is also included in determining its vertical structure  \citep[see Fig 6 in][]{ChandaP1}.

A surprising result is that the vertical constraining or pinching effect is felt over a large radial distance $\sim 500$ pc from the complex centre.  
This is due to the extended mass distribution in a complex. The gravitational force due to a spheroidal complex of constant density is maximum at its centre. Hence the constraining effect of the complex normal to the mid-plane is maximum at the centre of the complex and tapers off gradually with radial distance from the centre. The \HI scale height shows a significant decrease of a factor $\sim$ 3 at the centre of the complex, while the decrease is smaller but still $\sim 10 \%$, even at 500 pc.
The vertical scale height of stars and \HI would therefore appear locally corrugated over a scale of a few 100 pc around a complex. The \HI and stellar scale heights would  have a broad trough-like appearance with respect to the 
centre of the complex, unlike a uniform scale height distribution in an undisturbed disc.

A similar vertical constraining effect has been shown to apply to 
the general disc distribution at any point
in the Milky Way \citep{2002A&A...390L..35N,2018A&A...617A.142S} with a focus on the inner and the outer Galaxy, respectively. In these papers, a self-consistent vertical distribution has been obtained theoretically for a realistic multi-component disc consisting of gravitationally coupled stars, interstellar \HI and H$_2$ gas in the gravitational field of the dark matter halo. Due to the additional gravitational force of the other disc components and the halo, each disc component is shown to be constrained closer to the mid-plane, and the stellar density profile is shown to be steeper than sech$^2$. Thus the resulting scale height values at any radius are smaller than in a single-component disc. 
We stress that in contrast to the case of the complex where the additional force due to the complex on the disc is localized around a complex in the disc, here the gravitational force of the disc components and the halo 
is not localized to a given region. Rather the net vertical distribution of all disc components in the coupled disc is constrained (w.r.t their one-component values) at each radius.
Results from this model are shown to agree well with observations.  In particular, the resulting \HI gas scale height values as a function of radius for the inner Galaxy agree well with observations as shown by \citet{2002A&A...390L..35N}.
However, the local corrugation of \HI scale heights due to a molecular cloud complex as predicted by \citet{ChandaP1} has not been observationally confirmed so far.

\begin{figure*}
\begin{subfigure}{}
\includegraphics[width=1.\textwidth]{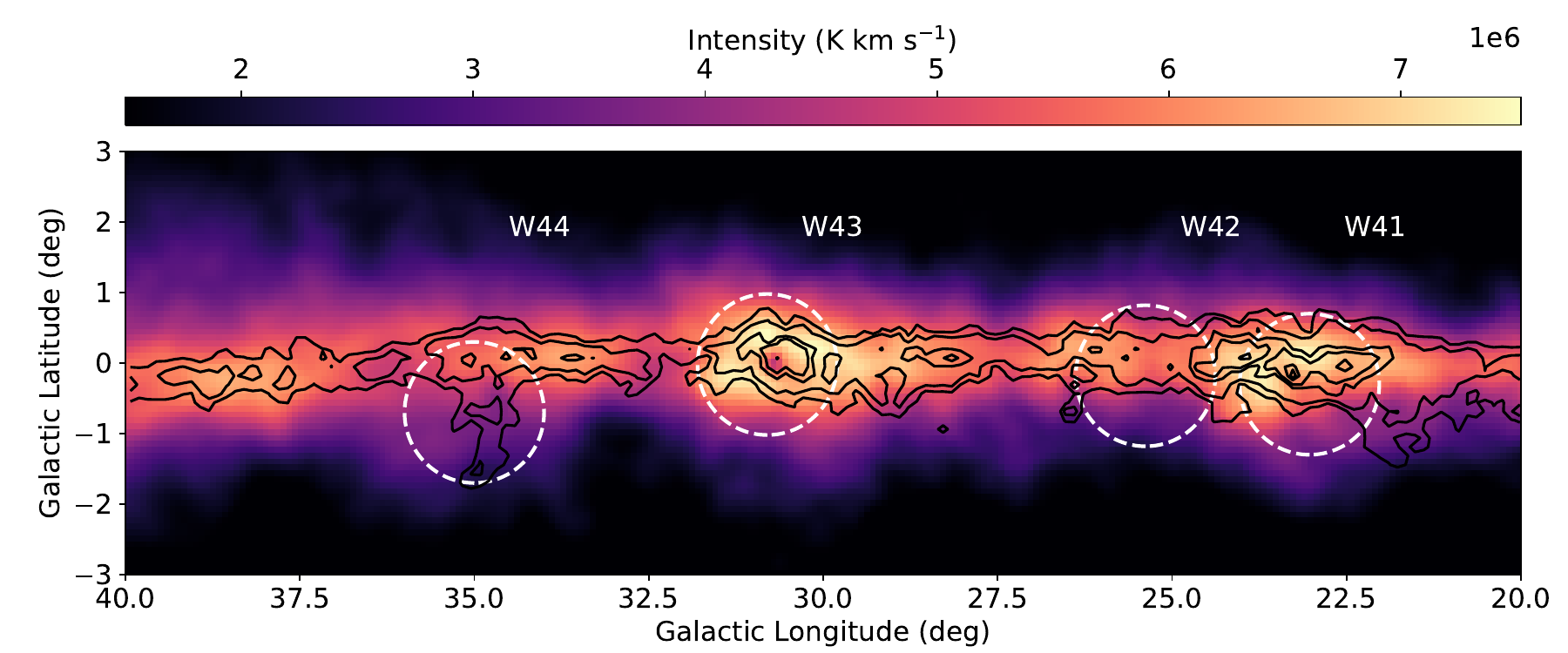}
\caption{The \HI 21cm line intensity distribution of the W41--W44 region integrated over the LSR velocity range $30-120 \kms$ is shown in colours. Overlaid black contours represent  the  $\12CO$ $1-0$ intensity integrated over the same velocity range. White dotted circles represent the positions and areas from which the \HI and CO spectra are extracted for each cloud. Values of CO contours are $\rm 2, 3, 5, 6, 8, 9, 10 \times 30 \ K \kms$.   }
\label{fig:Fig1}
\end{subfigure}

\begin{subfigure}{}
\includegraphics[width=1.\textwidth]{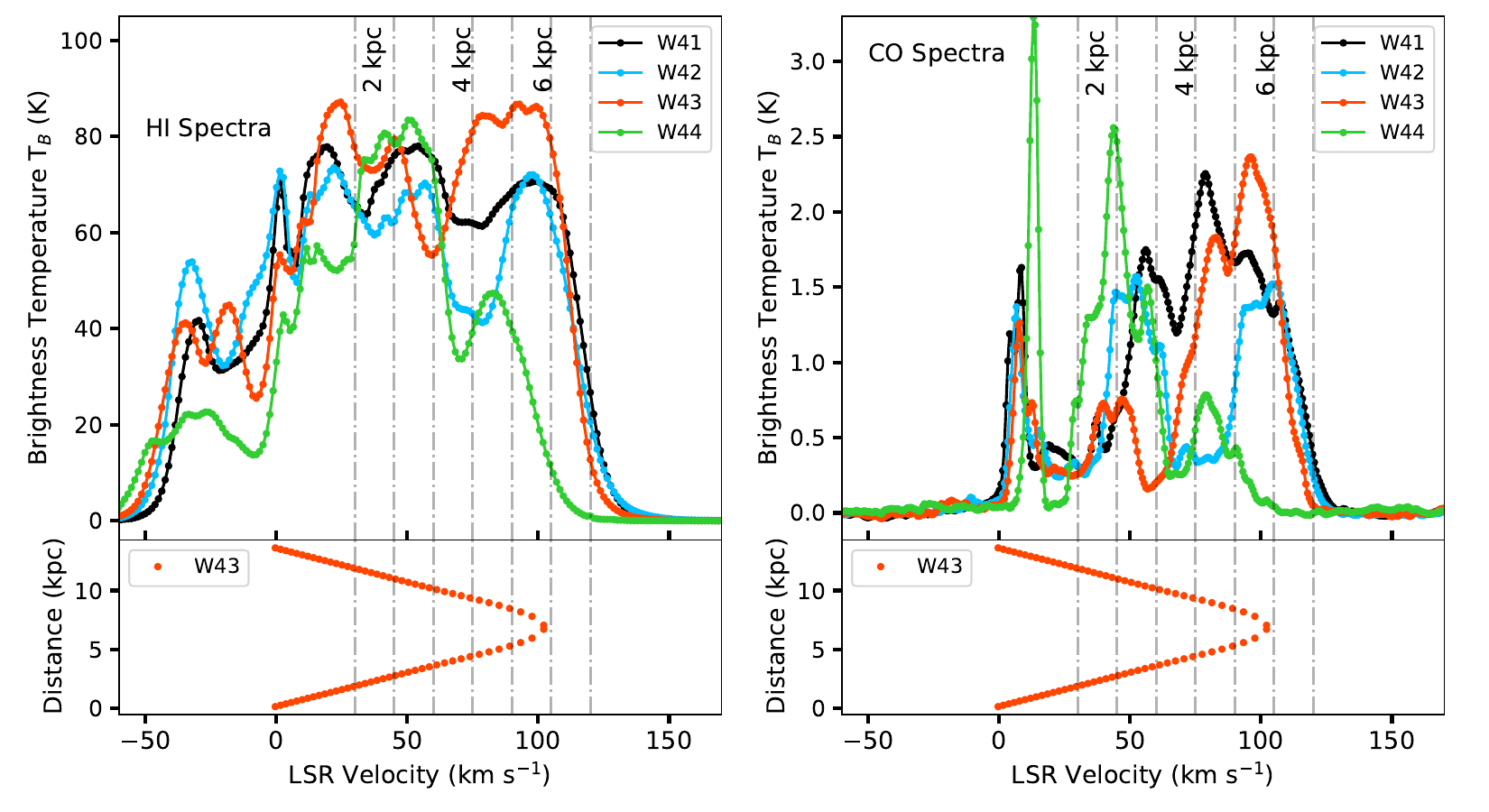}
\caption{Left and right top panels show \HI  21 cm and $\12CO$ $1-0$ spectra, respectively, extracted for circular areas of radius $1^{\circ}$ marked in figure~\ref{fig:Fig1}. The dot-dashed vertical lines demarcate the velocity bins discussed in the text. The bottom panels show the estimated line of sight distance against the LSR velocity at the position of W43 (i.e., $l = 30.8$)}
\label{fig:Fig2}
\end{subfigure}
\end{figure*}
\section{Sample and Observation}
\label{Sec:Data}

\subsection{The W41--W44 region}
\cite{1986ApJ...305..892D} identified 26 GMCs  with masses greater than $5 \times 10^5 \Msun$ in the inner Galaxy (i.e., within $l = -12^{\circ}$ and  $60^{\circ}$). In the Galactic longitude range encompassing
 W41, W42, W43 and W44 ($20^{\circ} < l <  40^{\circ}$) they found around 17  bright CO sources in their survey. A total of 11 of these  GMCs have masses greater than $10^5 \Msun$ and sizes ranging from $\sim60$ pc to $120$ pc (see their Table 2). In the following, we discuss the properties of these star-forming regions and  SNRs in the W41--W44  regions and their associated molecular cloud complexes.
W43 is a so-called mini-starburst region that is associated with one of the most massive GMC complexes in the Galaxy, which has a total H$_2$ gas mass of $\sim 7.1 \times 10^6 \Msun$ \citep{2011A&A...529A..41N,2023ApJS..264...16P}. \cite{2011A&A...529A..41N} conducted an extensive multi-wavelength study of the W43 region and identified  a large coherent complex of molecular clouds within an approximate diameter of $\sim 140$ pc. The overall mass contained in the high-density molecular clumps, possible sites of star formation, traced by $870 \ \mu$m dust emission in W43 is around $\sim 8.4 \times 10^5 \Msun$. 
Among the more than 20 clouds in this complex \citep{1987ApJ...319..730S, 2009ApJS..182..131R}, W43-Main ($l= 30.8^\circ$) and W43-South ($l= 29.9^\circ$) are the most massive and host most of the denser clumps 
\citep{2023ApJS..264...16P}. \cite{2011A&A...529A..41N} found that the main LSR velocity range of gas that is tightly associated with W43 ranges from $80$ to $110 \kms$. 
They did a Gaussian fitting in this velocity range to the $^{13}$CO spectrum integrated over  $1.8^{\circ}\times 0.8^{\circ}$ area that covers the entire W43, giving a peak at $\sim 95.9 \kms $  and FWHM of $\sim 22.3 \kms$. 
 The line of sight distance to the complex has been estimated to  be $6$ kpc, which is placed near the intersection of the Scutum Arm and the Galactic bar \citep{2011A&A...529A..41N,2014ApJ...781...89Z}.

The GMC G23.3$-$0.3, having a mass of $2 \times 10^6 \Msun$, host massive star formation 
regions in W41--W44 region and its distance has been found to be 4--5 kpc \citep{1986ApJ...305..892D,2006ApJ...643L..53A,2010ApJ...708.1241M,2014A&A...569A..20M}. The dominant CO emission towards this region is observed in the velocity range of $\sim 70-82 \kms$ \citep{2014A&A...569A..20M}.  Using the Galactic rotation curve of \cite{Reid2019}, the kinematic distance obtained from these velocities is $4.6$ kpc to $5.1$ kpc.
\cite{2015ApJ...811..134S} used their  $\12CO$, $^{13}\rm CO$ and $\rm C^{18}O$ data to identify  a large (i.e., $\sim 84 \times 15$ pc) and dense (i.e., $\sim 10^3 \rm cm^{-3}$ ) molecular complex with $\rm V_{LSR} = 77 \kms$ and distance $4.4$ kpc.
Within this region, a few SNRs remnants have been identified, including W41 ($l \sim 23.03 ^\circ$, $b \sim -0.3 ^\circ$ ) which is tightly associated with the GMC \citep{2008AJ....135..167L}. The presence of numerous stellar clusters and SNRs suggest ongoing massive star formation in this region \citep{ 2006AJ....131.2525H, 2008AJ....136.1477L, 2009BASI...37...45G,2010ApJ...708.1241M}.

The SNR W44, located at $l,b = 34.7^{\circ}, -0.4^{\circ}$ at distance of $\sim 3.2$~ kpc, is surrounded by a few GMCs whose CO emission peaks at $\sim 45 \kms$\citep{1998ApJ...505..286S,2004AJ....127.1098S, 2014A&A...565A..74C}. A GMC complex  
at this location  has been found to have  a mass of $1.8\times 10^6 \Msun$, a size $52$ pc and $\rm V_{LSR} = 44 \kms$ \citep{1986ApJ...305..892D}. An \HII region 
 G$034.8-0.7$ is also identified in the vicinity of the SNR W44. Various millimetre and infra-red observations of this environment  have shown that parts of the associated molecular clouds are shocked by the propagation of supernovae blast wave \citep{2005ApJ...618..297R,2009A&A...498..445P,2013ApJ...774...10S}. 

W42 is an obscured giant \HII region, characterized as a core-halo structure with an approximate size of $\sim 4.3$ pc located at $l = 25.38^{\circ}$ and $b = -0.18^{\circ} $ \citep{1985A&A...147...84W, 1993ApJ...418..368G, 2000AJ....119.1860B,2010A&A...523A...6D,2023ApJS..264...16P}. $^{13} \rm CO $  emission at  LSR velocities of $58 - 69 \kms$ is ascribed to a cloud physically associated to the W42  \HII region \citep{2009ApJS..181..255A,2013RAA....13..935A}. The distance to W42 has been estimated as $\sim3.8$ kpc \citep{2009ApJS..181..255A}. Various infrared images that trace the warm dust emission towards W42 reveal a bipolar nebular morphology of size $\sim 10$ pc \citep{2010A&A...523A...6D}.
\subsection{Observations and data preparation }
 To observationally probe the constraining effect, one needs to investigate the variation in the vertical distribution of \HI gas surrounding the molecular complexes in the W41--W44 region. To trace the atomic hydrogen content, we use \HI 21cm line emission, which has been extensively observed with high sensitivity in different parts of the Milky Way in multiple \HI surveys \citep[e.g.,][]{2005KalberlaLAB,2006StilVGPS,2016A&A...594A.116H}. 
Among the data for various CO rotational transition lines that trace the molecular hydrogen content in the clouds at different physical conditions, we use those of the $\12CO$ $J = 1-0$ line in our analysis whose, observed sky brightness commonly taken as a proxy to the molecular hydrogen content in the molecular ISM \citep{2013Bolatto}. Being the lowest rotational transition, the $\12CO$ line also traces the diffuse cloud structures contained in the complexes.
We obtained the publicly available data from two Galactic surveys that trace the \HI 21cm and $\12CO$ emission in the W41--W44 region, which are described in the following subsections. 
\subsubsection*{\HI data}
 To study the neutral hydrogen distribution in the W41--W44 region, we use the archival data from the all-sky database of the Galactic \HI $4\pi$ survey \citep[HI4PI,][]{2016A&A...594A.116H}. This comprises \HI 21cm observations from two surveys, the Effelsberg-Bonn \HI Survey  \citep[EBHIS,][]{2011AN....332..637K, 2016A&A...585A..41W} and the Galactic All-Sky Survey \citep[GASS,][]{2009ApJS..181..398M, 2010A&A...521A..17K,2015A&A...578A..78K}, both of which share similar sensitivity and angular resolution. The combined HI4PI data covers a radial velocity range from $-600 \kms$ to $600 \kms$ with a spectral resolution of $ 1.49 \kms $. The angular resolution of $16.2'$ and a noise level of $\sim 43$ mK, make  HI4PI the most sensitive \HI survey that covers the full sky.

\subsubsection*{CO data} \cite{2001ApJ...547..792D} released a complete survey in the $\12CO$ $J = 1-0$  of the entire Milky Way by combining their previous large-scale CO surveys. We use the $\12CO$ data taken from the First quadrant as part of this survey, which was observed with 
the CfA 1.2 m telescope. The data has a velocity coverage ranging from $-0.5$ to $271 \kms$ with a spectral resolution of $0.65 \kms$ and an angular resolution of $7.5'$. The data have a sensitivity of $0.18$ K covering the sky in Galactic latitudes $|b| < 32^{\circ}$ which is adequate for probing the vertical structures in our analysis. 
\begin{table}
\begin{tabular}{|c|c|c|c|}
\hline
        & \begin{tabular}[c]{@{}c@{}}Velocity range\\ \\ ($\kms$)\end{tabular} & \begin{tabular}[c]{@{}c@{}}Line of Sight\\ Distance\\ (kpc)\end{tabular} & \begin{tabular}[c]{@{}c@{}}Region with \\ peak $\12CO$ \\ emission\end{tabular} \\ \hline
Slice 1 & $30 - 45$                                                            & $2.32$                                                                    & W44                                                                             \\ \hline
Slice 2 & $45 - 60$                                                            & $3.15$                                                                    & W42,W44                                                                             \\ \hline
Slice 3 & $60 - 75$                                                            & $3.93$                                                                    & -                                                                               \\ \hline
Slice 4 & $75 - 90$                                                            & $4.87$                                                                    & W41,W43                                                                         \\ \hline
Slice 5 & $90 - 105$                                                           & $5.65$                                                                    & W41,W43                                                                             \\ \hline
Slice 6 & $105 - 120$                                                          & $6.93$                                                                    & W41,W43                                                                               \\ \hline
\end{tabular}
\caption{The velocity range of each slice and respective line of sight distance (near) estimated are tabulated here. We also mentioned the respective complexes which is emitting $\12CO$ significantly in each velocity range.}
\label{Tab1}
\end{table}

\subsubsection*{Velocity integrated intensity maps}
Figure~\ref{fig:Fig1} shows the \HI intensity distribution of the W41--W44 region integrated over the velocity range $30-120 \kms$. Overlaid black contours represent the integrated $\12CO$ intensity map in the same velocity range. White dotted circles marked the area from which the representative \HI and $\12CO $ spectra are extracted, which are given in  figure~\ref{fig:Fig2}. 
Figure~\ref{fig:Fig1} shows the integrated \HI and CO integrated emission, which comprises emission from gas clouds located at multiple positions along the line of sight. %
Additionally, the individual GMC complexes are emitting CO emission at different LSR velocities, indicating that they are located at different line of sight distances. 
Hence to investigate the constraining effect imposed by the individual molecular cloud complexes on their surrounding gas and also to minimize the uncertainty introduced in the  distance estimation, we did the following.
The position-position-velocity cubes obtained from the HI4PI and CO surveys have been integrated along the velocity axis from $30$ to $120 \kms$ in multiple bins, each having a width of $15 \kms$. 
The dot-dashed vertical lines in figure~\ref{fig:Fig2} demarcate each velocity bin. Hence each intensity map essentially represents a two-dimensional ``slice'', normal to the line of sight direction towards the W41--W44 region. Table~\ref{Tab1} shows the velocity range for each slice and the corresponding line of sight distance. Appendix~\ref{App_Dis} describes the estimation of  distance from observed LSR velocity. The bottom panel of figure~\ref{fig:Fig2} shows the estimated line of sight distance as a function of LSR velocity at the position of W43 ($l=30.8^{\circ}$). Due to the kinematic distance ambiguity, it is difficult to separate the emission from gas located at the near and the far distances, which both contribute to emission in the same velocity range. 
Because the angular scale of the structures representing far emission is smaller than that coming from a closer distance, the observed thickness of the disc will be better determined by emission from a distance nearer to us. Hence in this analysis, we consider the near distance as the line of sight distance to each slice. We also note that this uncertainty is minimal at $\rm V_{LSR} \gtrsim 90 \kms$ for the W41--W44 region. Unlike in other Galactic regions where the kinematic distance varies drastically with the Galactic longitude at the same LSR velocity, in the W41--W44 region the near kinematic distance remains the same (See figure~\ref{fig:Dist}). In other words, for a particular slice, the variation of the line of sight distance along the Galactic longitude is small enough so that we can assume it to be a constant for the entire W41--W44 region. Hence the near line of sight distance 
is given in table~\ref{Tab1}. Due to the geometry, the near kinematic distances increase in a linear fashion with $\rm V_{LSR}$ towards the W41--W44 direction and each slice is on an average $\sim 870$ pc distance apart.   We estimate the  distribution of \HI disc thickness in different slices and the following section discusses the estimation in detail.
\section{Scale height estimation}
\label{Sec:Est}
 
 The vertical density distribution of a single component isothermal disc in hydrostatic equilibrium follows a sech$^2$ profile \citep{1942ApJ....95..329S}. In addition to the sech$^2$ profile, exponential and Gaussian profiles are also used in much of the literature to model the vertical distribution  of the stars and gas. The scale height is defined as the HWHM of the vertical profile. 
 The density or mass-weighted rms heights (the second moment of the density distribution) are also considered as vertical height estimation in the literature \citep{2009ApJ...693.1346K,2012Hill}.  In the following, we give the details of the approach that we choose to estimate the thickness of the disc for a particular slice.

Each slice represents a density distribution of gas, $\rho(l,b)$, at the Galactic coordinates $l$ and $b$ which is located at a distance D away from us. At each Galactic longitude $l$, we take the density distribution along the latitude $b$ and use it to determine the disc thickness. We model the distribution by a Gaussian profile defined as follows,

\begin{equation}
\label{EqGauss}
    \rho(z) = \rho_0 \ \exp\left(\frac{-(z-z_0)^2}{2\sigma_{\rm Gauss}^2}\right),
\end{equation}
where the FWHM is given by  $\sqrt{8\ln 2}$ times the standard deviation, $\sigma_{\rm Gauss}$.  Here $z = \rm D\tan(b) $ is the height above the midplane in units of parsec. $z_0$ is the position at which the distribution is at its peak density, $\rho_0$.  The FWHM calculated from the best fitting $\sigma_{\rm Gauss}$ is considered as the estimate of \HI disc thickness in this analysis. Note that the estimated FWHM is double the \HI scale height of the disc. Therefore, any undulations in the \HI scale height, if present, will be readily detectable in the distribution of FWHM values. We also estimate the density-weighted rms of the height, $ \sigma_h$, as follows, 

\begin{equation}
\label{EqMom2}
    \sigma_h^2 =\frac{\int dz \ (z-\bar{z})^2 \rho(z) }{\int dz  \ \rho(z)},
\end{equation}
where $\bar{z}$ is the density-weighted mean position of the distribution. Though this is not a reliable quantity to directly compare with the scale height estimation from observations, it characterises the thickness of the disc with less bias than modelling it with a function. Thus estimated $ \sigma_h$ from eq~{\ref{EqMom2}} multiplied by a factor of $\sqrt{8\ln 2}$ can be considered as a proxy to the disc thickness.
\begin{figure*}
\label{fig:Fig3}
\includegraphics[width=.9\textwidth]{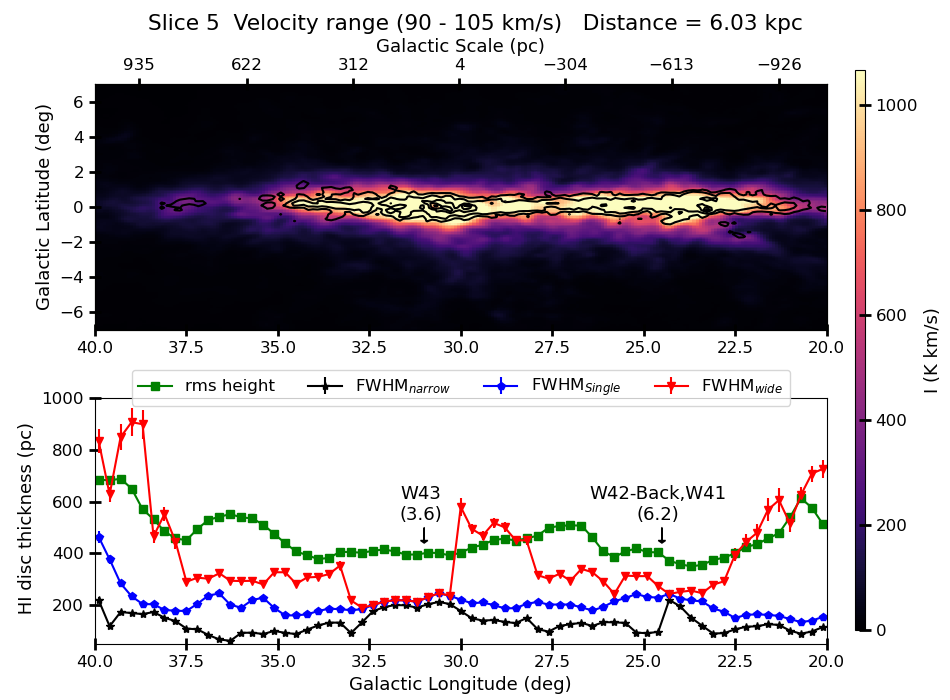}
\caption{The top panel shows the \HI intensity map integrated over the velocity range in Slice 5
and the respective CO intensity map overlaid as black contours. The $\12CO$ contours represent $\rm 0.6, 2, 5, 7, 9 \times 10.5 \ K \kms$. The bottom panel shows the \HI disc thickness estimated using different methods. FWHM values from the single Gaussian fitting and rms height are plotted as  blue pentagons and green squares, respectively. The rms height plotted here is $\sigma_{h}$ multiplied by $\sqrt{8\ln 2}$, so that it act as proxy to the overall disc thickness. FWHM of the narrow and wide components from double Gaussian fitting are given as black stars and red down triangles, respectively. All the FWHM values are plotted with error bars, which represent the fitting uncertainties. Two trough-like features in the FWHM$_{\rm wide}$ distribution, annotated by an arrow, are found at the location of  W43 and an extended structure contributed by W41 and a molecular cloud located at the back of the W42, (hence `W42--Back'). The total  H$_2$ mass obtained from the total $\12CO$ emission in the region of the trough is mentioned in parentheses in units of $10^6 \ \Msun$. }
\end{figure*}
We fit the vertical distribution $\rho(z)$ by a single Gaussian as given in eq.~\ref{EqGauss} and obtained the best fitting parameters $z_0,\rho_0$ and $\sigma_{\rm Gauss}$ for each value of $l$.  Hence the estimated FWHM from $\sigma_{\rm Gauss}$ (blue line with pentagon marker) with error bars as a function of Galactic longitude $l$ is given in figure~\ref{fig:Fig3} for ``Slice 5''. The error bar represents the uncertainty  of the fitting parameter estimates.
The top panel shows the \HI intensity map integrated over the velocity range in Slice 5
and the black contours are the respective integrated CO intensities. Assuming $l=30^{\circ}$ as the origin, we converted the Galactic longitudes to physical distance in units of parsec. These values are mentioned on the top of the x-axis.  The green line with square markers shows the density-weighted rms height for the \HI disc at each $l$.

In the $90-105 \kms$ velocity range,  the W43 complex dominates in $\12CO$ emission, while the GMC associated with W41  also contribute a significant amount of emission. In addition, a high velocity component towards the direction W42 is significantly contributing in this slice (See right panel of figure~\ref{fig:Fig2}). 
However, the molecular clouds contributing to this velocity component are not 
physically connected with the \HII regions of W41 and they are located behind W42, around a distance of $3$ kpc. We call this molecular cloud, which has significant $\12CO$ emission in slices 5 and 6, ``W42-Back''. From the top panel of figure~\ref{fig:Fig3}, we see that W43 has as  well compact high-density $\rm H_2$ structures  between $l= 29^{\circ}$ and $31^{\circ}$. However, from the best fitting FWHM of the single Gaussian profile, we do not see any hint of a decrease in the best fitting FWHM, which manifests the constraining effect.
Interestingly, an apparent increase in the disc thickness is visible around the complex centre (i.e., at $ l=30.6^{\circ}$). Similarly, around the GMCs W41 and W42--Back, we do not see any hint of constraining effect from the  best fitting FWHM. However,  \HI rms height distribution (green squares) shows interesting features around GMCs. The rms height variation around W43 shows a $\sim 700$ pc wide trough-like feature ($l=27.5^{\circ} - 35^{\circ}$). Similarly, around the location of W41 and W42--Back, we see another trough which is of similar size between $l= 21^{\circ}$ and $ 27^{\circ}$. This is clearly manifesting the gravitational pull that these complexes exert on the surrounding \HI gas, as discussed in \cite{ChandaP1}.
The distribution of rms height in the entire W41--W44 complex eventually shows that the overall thickness of the \HI disc  ranges from $\sim 400$ pc to $700$ pc. In contrast, the FWHM from the single Gaussian fitting varies from $\sim 150 $ pc to $250$ pc in this slice. These findings indicate the presence of an extended diffuse \HI disc with lower density compared to the gas residing near the mid-plane.

 The concept of vertical stratification within the atomic hydrogen disc has been known for long time and many attempts have led to the identification of multiple components (usually two) with discrete \HI scale heights \citep{1972ApJS...24...49R, 1975ApJ...198..281B}. By fitting the Galactic \HI vertical distribution  within Galactocentric radii of $0.4 R_{\sun} \lesssim R \lesssim R_{\sun}$ with a combination of two Gaussians and an exponential tail, \cite{1990ARA&A..28..215D} derived the thickness of \HI components.  
 The Gaussians have FWHM values  of $212$ pc and $ 530$ pc, where the narrow component has a peak density $\sim4$ times that of the wider component. The exponential tail has a thickness of $403$ pc with a peak density equal to one-sixth of that of the narrow component. Since the GMC complexes in the W41--W44 regions are also located within the  same range of Galactocentric radii, these results strongly suggest
the necessity of modelling the vertical distribution by at least a double Gaussian profile. Hence we repeat the exercise by fitting a double Gaussian defined as follows,

\begin{equation}
\label{EqGauss_d}
    \rho(z) = \rho^n_0 \ \exp\left(\frac{-(z-z^n_0)^2}{2\sigma_{\rm narrow}^2}\right) + \rho^w_0 \ \exp\left(\frac{-(z-z^w_0)^2}{2\sigma_{\rm wide}^2}\right),
\end{equation}
where the FWHM of each component can be obtained by multiplying $\sqrt{8\ln 2}$ to the respective $\sigma$s. 
  Various numerical simulations demonstrated that the vertical distribution of ISM is heavily influenced by dynamical processes such as turbulent mixing, stellar feedback etc \citep{2017Kritsuk,2023Chen}. MHD simulations of a vertical column of magnetised ISM by \citet{2012Hill} shows that the supernovae are the primary dynamical drive that leads to vertical stratification of three-phase ISM, where gas components at different temperature regime have distinct scale heights. 
However, separating the three phases through observations remains challenging due to several factors reasons including difficulties in temperature measurements, velocity crowding and components overlapping in velocity etc \citep{Bhattacharjee2023}.
Hence, in this analysis, we only focus to distinguish between \HI close to the disc and \HI distributed over a thicker region, by modelling the vertical distribution by the double Gaussian profile. The two components of the disc obtained from the fitting procedure may not necessarily  represent any distinct gas phases.

We use the parameters from the single Gaussian fitting and the rms height estimates to feed the initial guesses while modelling double Gaussian profile.
It is possible that for some region, two Gaussian are not required and a single Gaussian model is adequate to fit the observational data. In such cases, the wider Gaussian profile will be considering the values lying within the noise of the data for modelling and the resultant peak density, $\rho^w_0$, will be only a few times the noise floor. Additionally, in such cases, the narrow Gaussian component is expected to have fitting parameters similar to those of a single Gaussian. Based on these criteria, we performed fitting with both single and double Gaussian models for the vertical distribution at different longitudes. Wherever the wider component's fitting parameters (especially $\rho^w_0$) are insignificant with respect to the noise floor and the fitted values of narrow components ($\rho^n_0$ and $\sigma_{\rm narrow}$) match 
those of a single Gaussian within  85\%, we assume that the disc can be modelled by a single Gaussian profile and have a single scale height at that longitude. The  thus estimated FWHM of \HI narrow and wider Gaussian components at different longitudes are represented as red and black points, respectively, with error bars in figure~\ref{fig:Fig3}. 
Here, the error bars correspond to the uncertainty  in the fitting estimates. Note that, at values of $l$ for which a single Gaussian represents the vertical distribution better than the double Gaussian profile, the FWHM of the wider component is replaced with that of the fitted single Gaussian. Such $l$ values can be easily identified from figure{~\ref{fig:Fig3}}; the FWHM of the single Gaussian and wider components match at those locations.  
Interestingly, in this figure{~\ref{fig:Fig3}}, such positions are highly 
correlated with the locations of the  W41 and W43 GMC complexes. This indicates the lack of a significant thick diffuse component of atomic hydrogen gas around these complexes. In other words, gas located nearer to the complex is getting pulled towards the mid-plane by its gravitational potential. As we move farther away from the complex, the effect of the potential on diffuse gas will diminish,
leaving it as an extended halo. This eventually indicates the pinching or constraining effect that \cite{ChandaP1} have found through theoretical modelling.

We determine  the different estimates of \HI disc thickness as a function of $l$ with a pixel size of $0.08^{\circ}$. To reduce noise and hence to make prominently visible the trough-like features, all the \HI disc thickness estimates have been smoothed by averaging in uniform bins. For the averaging, we have chosen a bin size of  $\sim 0.3^{\circ}$, which is similar to the angular resolution of the \HI data ($\sim 0.26^{\circ}$). The broad troughs between $l=21^\circ$ and $29^\circ$ and a narrow trough between $l=30^\circ$ and $33^\circ$ in the FWHM of \HI wide component  provide clear evidence for vertical constraining that the GMC complexes of W41, W42--Back and W43 respectively are imposing on the \HI disc. We annotated those locations where pinching is  visible in figure~\ref{fig:Fig3} by an arrow and the name of the respective GMC complex. The H$_2$ mass in units of $10^6 \ \Msun$ obtained from measuring the integrated intensity of $\12CO(1-0)$ emission in the region where the trough is visible is mentioned in figure~\ref{fig:Fig3} underneath the name of the respective complex. 
We use the conversion factor $2.75 \times 10^{20}$ $(\rm cm^2 \ K \ \kms)^{-1}$ given by \cite{Bloemen1986} to calculate the H$_2$ column density and further H$_2$ mass. We repeat this exercise for other slices mentioned in Table~\ref{Tab1} and plots similar to Fig~\ref{fig:Fig3} for other slices are given in Appendix~\ref{App_SH}. In the following section, we discuss about our findings on the pinching effect by various complex.

\section{Results}
\label{Sec:Res}

As shown in \cite{ChandaP1}, the gravitational effect of molecular complexes results in local corrugations in the thickness of stellar and gas disc. These are the additional variations on top of the global corrugation of the disc mid-plane. Our analysis shows strong evidence for the pinching effect in the W41--W44 region. The broad trough in the FWHM of the wide component (red points in figure~\ref{fig:Fig3}), between $l=21^{\circ}$ and $29^{\circ}$ is a strong manifestation of the gravitational constraint imposed by the molecular gas in  W41 and W42--Back.  In this velocity range ($90 - 105 \kms$), both W41 and W42--Back have significant $\12CO$ emission in the form of an elongated structure along the mid-plane with a few high-density clumps at the cloud centre (i.e., at $l =23.03 ^\circ$ and $l =25.38 ^\circ$). The location of a broad minimum (from $23^\circ$ to $26^\circ$) in the trough aligns with the position of these cloud centres. With respect to the thickness of the disc far away from the clouds, the FWHM of the wider component is reduced by a factor of 2.5 around the complex centre. The molecular hydrogen mass from total $\12CO$ emission comprising the entire region where the trough is visible is $6.4 \times 10^6 \Msun$. 
The width of the trough is around $\sim 850$ pc.  A similar trough that however is less wide ($\sim 470$ pc) is also visible around W41 in the next lower velocity range, i.e., Slice 4 ($75 - 90 \kms$) (See figure~\ref{fig:SH_Slice4}). 
Interestingly, in this velocity range the $\12CO$ spectrum of W42 doesn't show any significant emission while W41 has the strongest emission. Clearly, we can say that the pinching in thickness of the wider component seen in this slice is solely caused by the molecular cloud complex in W41. The similarity in the length of the dense elongated $\12CO$ structure in W41  ($\sim500$ pc) and the width of the trough is also strongly favouring the evidence for pinching.
It is to be noted that  similar trough-like features are also visible in the rms-height distribution around W41 in both slices. As we go to the other slices where the LSR velocity decreases or the slices become nearer to us, the trough-like feature is gradually vanishing. This is clearly because the W41 does not have significant molecular hydrogen content at those distances. At high velocity ranges $\rm 105 < V_{LSR} < 120 \kms$ (i.e., Slice 6), W42--Back and W41 have moderate amount of emission which causes another trough in FWHM$\rm _{wide}$ and rms height in similar longitude range. The emission from molecular clouds associated with W42 is significant in Slice 2, however, a pronounced deviation in either FWHM$\rm _{wide}$ or rms height is not detected.

W43 is the most massive complex in the entire region under consideration; it has its peak CO emission in Slice 5. From figure~\ref{fig:Fig3}, we see a narrow dip near W43 from $l =30^\circ$ to $33^\circ$. The GMC complex in W43 is positively attracting the surrounding gas, which results in the existence of a single denser \HI structures confined around the mid-plane. The width of the trough is $\sim 320$ pc and the disc thickness is reduced by a factor of $\sim2$ at the minima. At Slice 3 and Slice 4, W43 have concentrated dense CO structures in a similar longitude range and a decrease in the FWHM$\rm _{wide}$ is visible from $l=29^{\circ}$ to $33^{\circ}$ in figure~\ref{fig:SH_Slice3} and \ref{fig:SH_Slice4} respectively. The average width of the trough is around $350$ pc in both cases and the denser structures are located near the centre of the trough. A similar narrow trough can be seen in Slice 6, though not as prominent as in other slices. Again, wherever the pinching effect by the W43 complex is detected from the FWHM$\rm _{wide}$ distribution, their respective rms height distribution also shows a coinciding trough-like feature, but lesser in magnitude. The $\12CO$ spectra towards the direction of W43 show a low velocity component from $30$ to $55 \kms$ in figure~\ref{fig:Fig2} and \cite{2011A&A...529A..41N} identified this to be originated from diffuse clouds located in the front along the line of sight, unrelated to W43 complex. 
In Slice 4, we see dense structures in $\12CO$ emission around $l= 30^{\circ}$, which are originating from these front clouds. The narrow dip in FWHM$\rm _{wide}$ from $l = 29^{\circ}$ to $32^{\circ}$ is caused by the constraining effect of these clouds, which we name ``W43-Front''. The total $H_2$ mass from this W43-Front is $2 \times 10^5 \ \Msun$ and the FWHM is decreased by a factor of 1.5. This finding strongly suggests the possibility of detecting the constraining effect of less massive molecular clouds or complexes,  where the trough in disc thickness is generally lower in magnitude.

Unlike the other cases, for W44, the effect of pinching is only detected in the velocity range $45 - 60 \ \kms$. The width of the trough detected in Slice 2 around W44 is about $165$ pc and the thickness of the diffuse \HI component decreases by a factor of three. The main velocity component of W44 is $30 - 60 \kms$, with the denser CO structures emitting around $45 \kms$. Although a trough feature is visible around W44 in the other velocity range $30 - 45 \kms$, a strong increase in the FWHM at the exact location of the W44 complex, prevents us from a confirmed detection for pinching in Slice 1. In contrast to the other complexes, which are confined to the mid-plane of the Galactic disc, W44 has a filament-like structure that extends along the vertical axis. 
The molecular gas associated with the SNRs W44 has been found to be 
strongly shocked \citep{2004AJ....127.1098S,2013ApJ...774...10S}. Indeed  $> 25 \kms$ wide high-velocity CO line wings 
suggest a violent interaction between the molecular cloud and the SNR. Additionally, ultra-high velocity emissions in CO J$=3-2$, traced by line widths $\sim 100 \ \kms$ are also detected toward W44 \citep{2013ApJ...774...10S}. 
Possibly such a violent environment has disrupted the entire gas distribution surrounding it. Eventually this results in making the surrounding vertical disc structure more complex and hence detecting constraining effect become more difficult. Table~\ref{Tab2} summarises the details about the detected troughs in our analysis. In addition to the GMC complexes of our interest, there are enormous amounts of $\12CO$ emission at other Galactic longitudes. For instance, in Slice 1, the longitude range from $25^{\circ}$ to $31^{\circ}$ comprises an elongated dense molecular gas structure having a total mass of $\sim2 \times 10^5 \ \Msun$. 
This cloud structure creates a trough in a similar longitude range, with a similar fractional decrease in thickness. Similarly, the narrow dips centred at $28^{\circ}$ and $37.5^{\circ}$ in Slice 4, are also possibly caused by the constraining effect of the dense gas clouds present in the respective location. Hence these local corrugations that are prominently visible in the thickness of \HI diffuse disc, are in very good agreement with the predictions by \cite{ChandaP1}.

    \begin{table}
\centering
\setlength{\tabcolsep}{1.1pt}
\renewcommand{\arraystretch}{1.1}
\begin{tabular}{|c|c|c|c|c|c|}
\hline
\begin{tabular}[c]{@{}c@{}}GMC\\ Complex\end{tabular}                                                 & \begin{tabular}[c]{@{}c@{}}Distance\\to trough\\  (kpc)\end{tabular} & \begin{tabular}[c]{@{}c@{}}Galactic \\ Longitude \\ range\end{tabular} & \begin{tabular}[c]{@{}c@{}}Mass\\ ($\times 10^6  \Msun$)\end{tabular} & \begin{tabular}[c]{@{}c@{}}Width \\ (pc)\end{tabular} & \begin{tabular}[c]{@{}c@{}}Height\\ (pc)\end{tabular} \\ \hline
W44                                                     & $3.15$                                                    & $32.5-35.5$                                                            & $0.5$                                                                 & $165$                                                 & $400$                                                 \\ \hline

W43--Front                                              & $3.15$                                                    & $29.0-32.0$                                                            & $0.2$                                                                 & $165$                                                 & $200$                                                 \\ \hline
W43                                                     & $4.10$                                                    & $27.5-32.5$                                                            & $1.1$                                                                 & $360$                                                 & $550$                                                 \\ \hline
W43                                                     & $4.87$                                                    & $29.0-33.0$                                                            & $1.3$                                                                 & $340$                                                 & $180$                                                 \\ \hline
W43                                                     & $6.03$                                                    & $30.0-33.0$                                                            & $3.6$                                                                 & $320$                                                 & $170$                                                 \\ \hline
W43                                                     & $6.93$                                                    & $29.0-32.5$                                                            & $1.0$                                                                 & $420$                                                 & $450$                                                 \\ \hline
W41                                                     & $4.10$                                                    & $21.5-24.5$                                                            & $1.3$                                                                 & $220$                                                 & $250$                                                 \\ \hline
W41                                                     & $4.87$                                                    & $20.0-25.5$                                                            & $2.6$                                                                 & $470$                                                 & $350$                                                 \\ \hline

\begin{tabular}[c]{@{}c@{}}W41,\\ W42--Back\end{tabular} & $6.03$                                                    & $21.0-29.0$                                                            & $6.4$                                                                 & $850$                                                 & $350$                                                 \\ \hline
\begin{tabular}[c]{@{}c@{}}W41,\\ W42--Back\end{tabular} & $6.03$                                                    & $22.5-26.0$                                                            & $3.4$                                                                 & $420$                                                 & $400$                                                 \\ \hline
\end{tabular}
\caption{The table summarises the details about the detected troughs in \HI FWHM$_{\rm wide}$ distribution.  }
\label{Tab2}
\end{table}
\begin{figure*}
\includegraphics[width=1\textwidth]{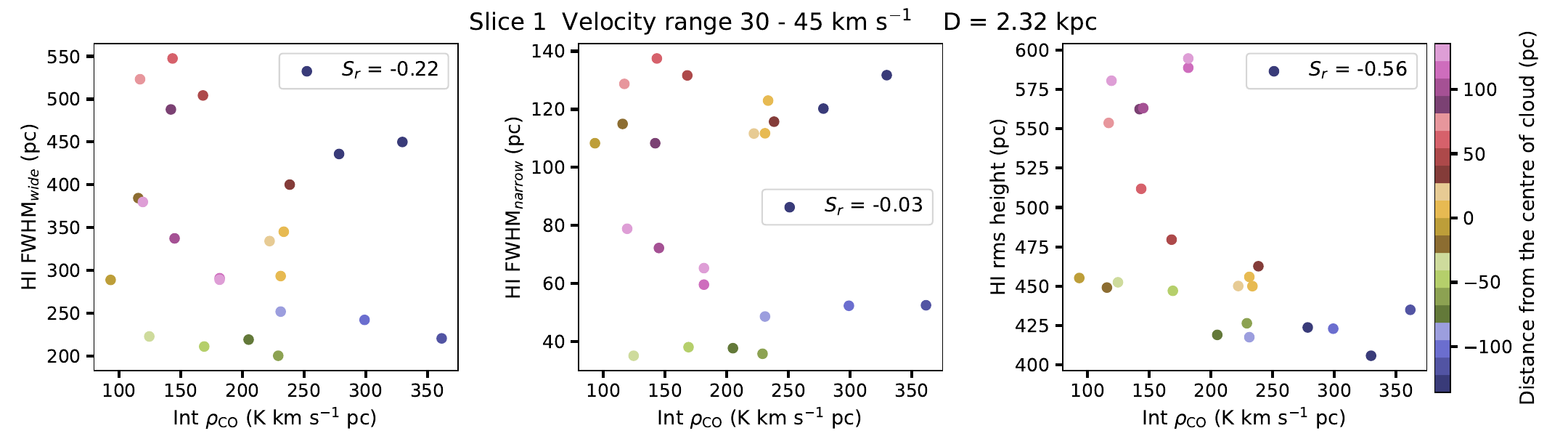}
\includegraphics[width=1\textwidth]{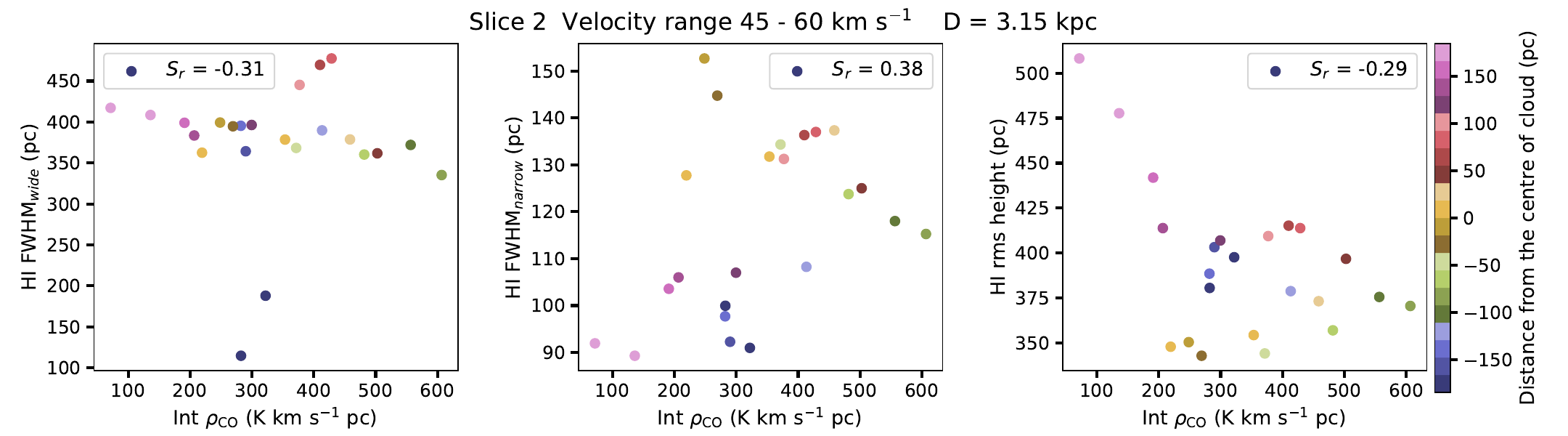}
\includegraphics[width=1\textwidth]{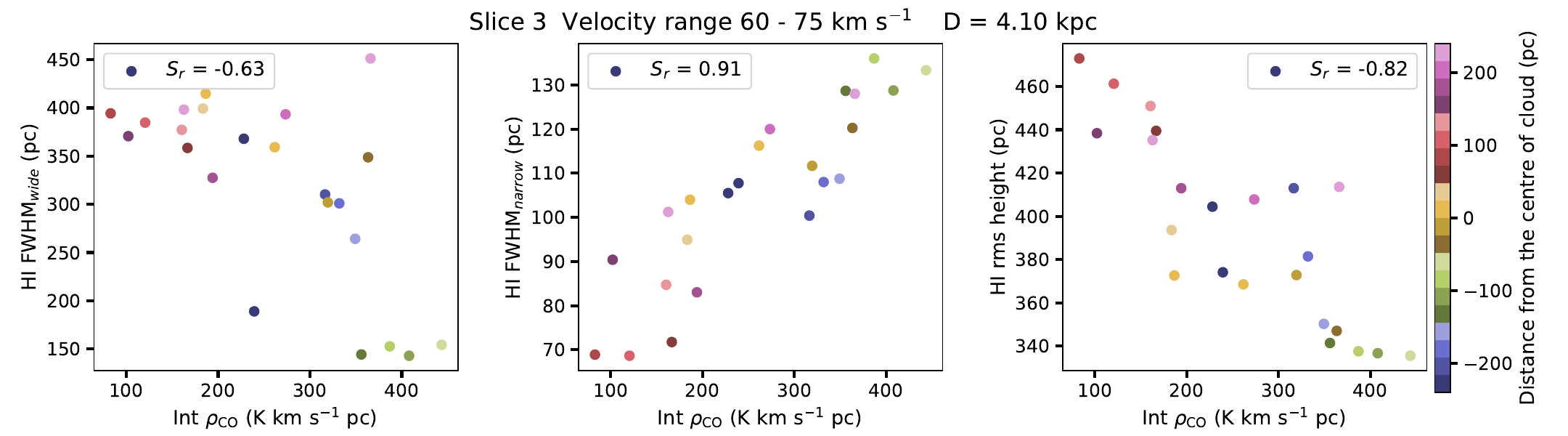}
\includegraphics[width=1\textwidth]{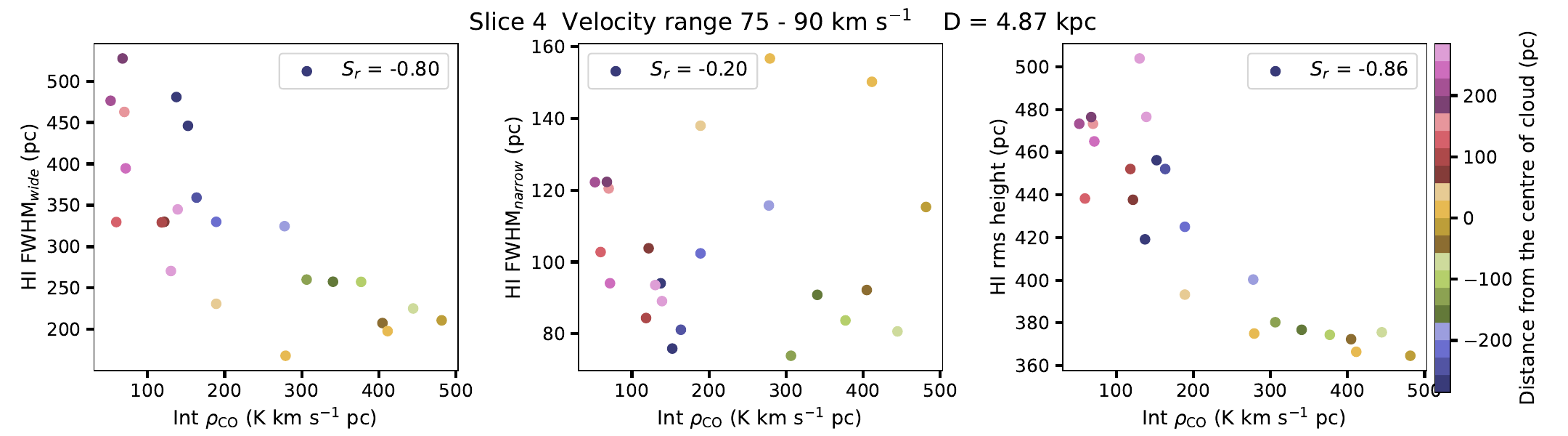}
\caption{Correlation study in W41-W42 region ($l= 20^{\circ}- 27^{\circ}$) at different slices. The scatter plots between the integrated $\rho_{CO}$ and FWHM$_{\rm wide}$ (left column), FWHM$_{\rm narrow}$ (middle column) and \HI rms height (right column). Each row corresponds to different slices. Points are colour-coded with distance from the centre of the cloud.}
\label{fig:FigCorr1}
\end{figure*}

\begin{figure*}
\includegraphics[width=1\textwidth]{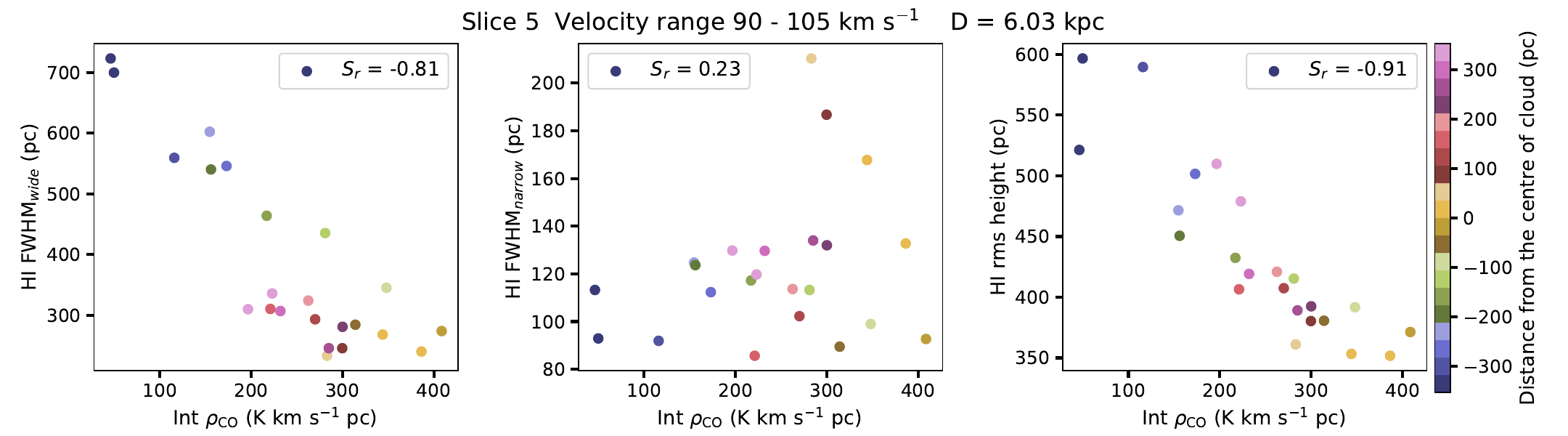}
\includegraphics[width=1\textwidth]{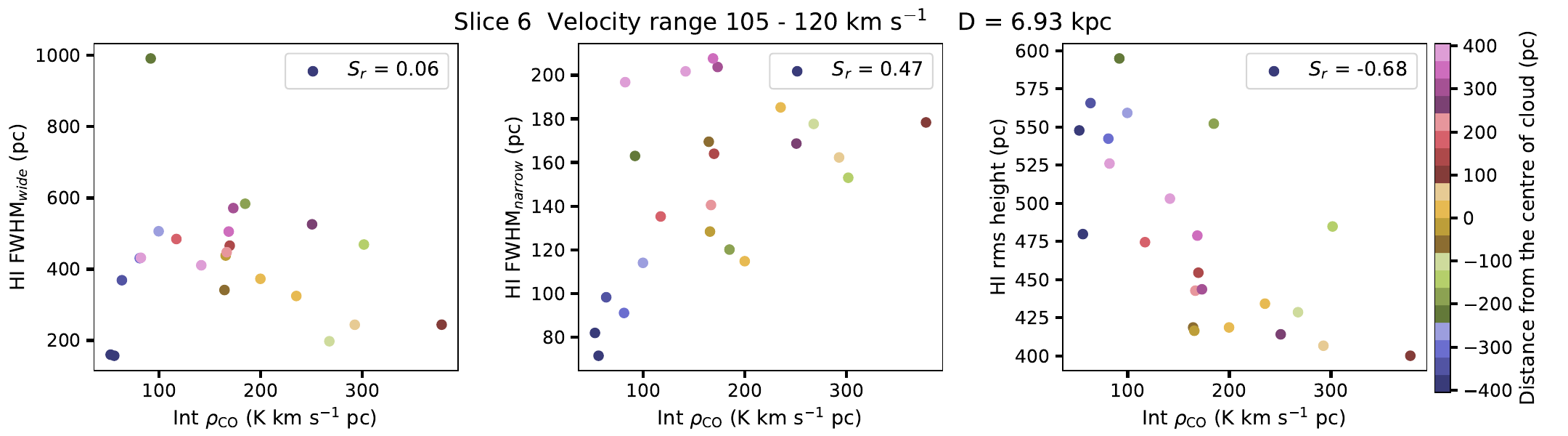}
\caption{Continuation of figure~\ref{fig:FigCorr1}}
\label{fig:FigCorr2}
\end{figure*}
 \begin{table*}
\setlength{\tabcolsep}{1pt}
\renewcommand{\arraystretch}{1.3}
\begin{tabular}{|c|ccccccccc|}
\hline
        & \multicolumn{9}{c|}{Spearman correlation coefficient ($S_r$)}                                                                                                                                                                                                                                                                                                                                                                                                                                                                                                                                                                                                                                                                                                                                                                                                                                                                                                                                                                     \\ 
        & \multicolumn{3}{c|}{W41- W42 ($l = 20 ^{\circ} \ - \ 27 ^{\circ} $ )}                                                                                                                                                                                                                                                                            & \multicolumn{3}{c|}{W43 ($l = 27.5 ^{\circ} \ - \ 32.5 ^{\circ} $ )}                                                                                                                                                                                                                                                                             & \multicolumn{3}{c|}{W44 ($l = 33 ^{\circ} \ - \ 37 ^{\circ}$ )}                                                                                                                                                                                                                                                             \\ \hline
        & \multicolumn{1}{c|}{\begin{tabular}[c]{@{}c@{}}$\rm FWHM_{wide}$\\ v/s\\ $\rm Int \ \rho_{CO}$\end{tabular}} & \multicolumn{1}{c|}{\begin{tabular}[c]{@{}c@{}}$\rm FWHM_{narrow}$\\ v/s\\ $\rm Int \ \rho_{CO}$\end{tabular}} & \multicolumn{1}{c|}{\begin{tabular}[c]{@{}c@{}}rms height\\ v/s\\ $\rm Int \ \rho_{CO}$\end{tabular}} & \multicolumn{1}{c|}{\begin{tabular}[c]{@{}c@{}}$\rm FWHM_{wide}$\\ v/s\\ $\rm Int \ \rho_{CO}$\end{tabular}} & \multicolumn{1}{c|}{\begin{tabular}[c]{@{}c@{}}$\rm FWHM_{narrow}$\\ v/s\\ $\rm Int \ \rho_{CO}$\end{tabular}} & \multicolumn{1}{c|}{\begin{tabular}[c]{@{}c@{}}rms height\\ v/s\\ $\rm Int \ \rho_{CO}$\end{tabular}} & \multicolumn{1}{c|}{\begin{tabular}[c]{@{}c@{}}$\rm FWHM_{wide}$\\ v/s\\ $\rm Int \ \rho_{CO}$\end{tabular}} & \multicolumn{1}{c|}{\begin{tabular}[c]{@{}c@{}}$\rm FWHM_{narrow}$\\ v/s\\ $\rm Int \ \rho_{CO}$\end{tabular}} & \begin{tabular}[c]{@{}c@{}}rms height\\ v/s\\ $\rm Int \ \rho_{CO}$\end{tabular} \\ \hline
Slice 1 & \multicolumn{1}{c|}{$-0.22$}                                                                                 & \multicolumn{1}{c|}{$-0.03$}                                                                                   & \multicolumn{1}{c|}{$-0.56$}                                                                                      & \multicolumn{1}{c|}{$-0.51$}                                                                                 & \multicolumn{1}{c|}{$-0.49$}                                                                                   & \multicolumn{1}{c|}{$0.31$}                                                                                     & \multicolumn{1}{c|}{$-0.29$}                                                                                  & \multicolumn{1}{c|}{$0.13$}                                                                                    & $-0.34$                                                                                      \\ \hline
Slice 2 & \multicolumn{1}{c|}{$-0.31$}                                                                                 & \multicolumn{1}{c|}{$0.38$}                                                                                    & \multicolumn{1}{c|}{$-0.29$}                                                                                     & \multicolumn{1}{c|}{$-0.33$}                                                                                 & \multicolumn{1}{c|}{$-0.56$}                                                                                   & \multicolumn{1}{c|}{$-0.07$}                                                                                     & \multicolumn{1}{c|}{$-0.09$}                                                                                 & \multicolumn{1}{c|}{$0.75$}                                                                                    & $-0.93$                                                                                      \\ \hline
Slice 3 & \multicolumn{1}{c|}{$\mathbf{-0.63}$}                                                                                 & \multicolumn{1}{c|}{$\mathbf{0.91}$}                                                                                    & \multicolumn{1}{c|}{$\mathbf{-0.82}$}                                                                                      & \multicolumn{1}{c|}{$-0.43$}                                                                                 & \multicolumn{1}{c|}{$-0.18$}                                                                                   & \multicolumn{1}{c|}{$-0.91$}                                                                                      & \multicolumn{1}{c|}{$-0.13$}                                                                                 & \multicolumn{1}{c|}{$0.55$}                                                                                    & $-0.92$                                                                                      \\ \hline
Slice 4 & \multicolumn{1}{c|}{$\mathbf{-0.80}$}                                                                                 & \multicolumn{1}{c|}{$-0.20$}                                                                                   & \multicolumn{1}{c|}{$\mathbf{-0.86}$}                                                                                      & \multicolumn{1}{c|}{$0.49$}                                                                                  & \multicolumn{1}{c|}{$-0.26$}                                                                                   & \multicolumn{1}{c|}{$-0.66$}                                                                                      & \multicolumn{1}{c|}{$-0.20$}                                                                                 & \multicolumn{1}{c|}{$0.34$}                                                                                    & $-0.65$                                                                                      \\ \hline
Slice 5 & \multicolumn{1}{c|}{$\mathbf{-0.81}$}                                                                                 & \multicolumn{1}{c|}{$0.23$}                                                                                    & \multicolumn{1}{c|}{$\mathbf{-0.91}$}                                                                                      & \multicolumn{1}{c|}{$0.21$}                                                                                  & \multicolumn{1}{c|}{$0.60$}                                                                                    & \multicolumn{1}{c|}{$-0.66$}                                                                                      & \multicolumn{1}{c|}{$-0.02$}                                                                                 & \multicolumn{1}{c|}{$0.47$}                                                                                   & $-0.90$                                                                                      \\ \hline
Slice 6 & \multicolumn{1}{c|}{$0.06$}                                                                                  & \multicolumn{1}{c|}{$0.47$}                                                                                    & \multicolumn{1}{c|}{$\mathbf{-0.68}$}                                                                                      & \multicolumn{1}{c|}{$-0.26$}                                                                                 & \multicolumn{1}{c|}{$-0.68$}                                                                                   & \multicolumn{1}{c|}{$-0.71$}                                                                                      & \multicolumn{1}{c|}{$-0.71$}                                                                                 & \multicolumn{1}{c|}{$-0.91$}                                                                                   & $-0.86$                                                                                      \\ \hline

\end{tabular}
\caption{Spearman correlation coefficient between the various quantities for individual complex regions. Each row corresponds to the different slices.}
\label{Tab3}
\end{table*}
To quantify the correlation between the \HI scale height variation and GMCs, we estimate the Spearman correlation coefficient between the disc thickness parameters and CO intensity. For every value of $l$, we estimate the integrated CO intensity summed over the entire vertical axis. For any slice, this integrated $ \ \rho_{CO}$ is proportional to the total molecular gas density summed over the entire vertical axis at a particular longitude over the respective velocity range.
Note that the angular resolution of the \HI and CO data are different and hence to have a position-wise comparison between the \HI disc thickness  and CO intensity, we did a uniform binning along the longitude axis with a bin width of $0.3^{\circ}$. Hence 
the bin averaged values of different quantities are considered for the correlation study. We separate the entire W41--W44 region into three regions so that the effect of the individual complex can be quantified. 
For example, we consider the Galactic longitudes from $l=20^{\circ} $ to $l=27^{\circ} $, where the W41 and W42--Back clouds are located.  
Figures~\ref{fig:FigCorr1} and \ref{fig:FigCorr2} show the scatter plot between the integrated $\rho_{CO}$ and FWHM$\rm _{wide}$ (left column), FWHM$\rm _{narrow}$ (middle column) and the \HI rms height (right column) for W41--W42 region. Each row corresponds to different slices. Points in the scatter plots are colour-coded with distance from the centre of the complex (here it is $ l=23.5^{\circ}$). The Spearman correlation coefficient between the respective quantities is mentioned in each plot.  The coefficient ranks the correlation between two quantities, where a value of $1$ corresponds to a strong correlation and $0$ corresponds to zero association between them. A Spearman correlation coefficient of $-1$ represents a strong  anti-correlation between two quantities. From figure~\ref{fig:FigCorr1} and figure~\ref{fig:FigCorr2}, we can see that FWHM of the wide component is  anti-correlated with the integrated CO emission in the velocity range $60 - 105 \kms$. Respective values of Spearman coefficient $-0.63$, $-0.80$ and $-0.81$
at different slices in this velocity range depict the following picture. In regions where the thickness of the \HI wide component is smaller, the molecular gas has a higher mass. In other words, high-density CO regions are reducing the thickness  of \HI wide component, which is nothing but the effect of pinching. In slices where anti-correlation is strongest (Slice 4 and Slice 5), we can easily see from the scatter plot of \HI FWHM$_{\rm wide}$ and  integrated $\rho_{CO}$, the points that are lying in the high-density tail is essentially located around the complex centre. 
The decreasing trend in the anti-correlation as it goes towards the lower velocity components suggests that the pinching effect is less due to lower molecular mass. Similarly, the \HI rms height shows a very strong anti-correlation with the integrated $\12CO$ intensity in similar velocity ranges, with Spearman coefficient $-0.82$, $-0.86$ and $-0.91$ in each slice.
The negative correlation of FWHM$_{\rm wide}$ and rms height with the H$_2$ density in these velocity ranges quantitatively confirms the detection of vertically constrained \HI disc by molecular complexes in  W41 and W42--Back. The Spearman coefficient of $-0.68$ between the \HI rms height and the integrated CO intensity in Slice 6, also favours the pinching effect by W41, W42--Back together.   We did a similar analysis in the other two complexes and their estimates of the Spearman correlation coefficients are given in table~\ref{Tab3}.

The middle column of figure~\ref{fig:FigCorr1} and figure~\ref{fig:FigCorr2} is quantitatively representing any association between the narrow disc thickness and molecular gas density, if it exit. 
Though any noticeable  trend of neither correlation nor anti-correlation is visible, the following are some exceptions.
The correlation between FWHM$_{\rm narrow}$ and integrated $\rm \rho_{CO}$ in W41--W42 region at velocity range $60-75 \kms$ is $0.9$. \cite{2008AJ....135..167L} identify a large molecular CO cloud at $77 \kms$ interacting with the supernova W41, at a kinematic distance $3.9-4.5$ kpc. The strong correlation  of $0.9$ between the molecular gas and narrow disc is probably the resultant effect of this interaction. Similarly, the surrounding gas around W41 at a velocity range of $45-60 \kms$ is found to be violently disrupted by supernovae shock, which could possibly result in a small hump-like feature seen in the thickness of narrow disc between $20^{\circ}$ to $25^{\circ}$ (Black stars in figure~\ref{fig:SH_Slice3}). The fractional increase in the thickness is around a factor of two. Similarly, at the location of W43 in Slice 5, where the complex has its brightest emission in $\12CO$, an apparent increase in the scale height of the narrow disc by two times, is visible from $30^{\circ}$ to $32^{\circ}$. A positive rank correlation with $S_r=0.6$  at the location of W43 in Slice 5 is a result of this variation in FWHM$_{\rm narrow}$. The dense \HI gas that is seen confined to the mid-plane within $\lesssim 200$ pc can easily get disrupted by the local dynamics happening within the high star-forming regions like W41 and W43 and thus could possibly enhance its thickness. 
A positive Spearman coefficient of $S_r=0.75$ between narrow disc thickness and CO intensity  around the W44 region in velocity slice 2, where the related SNR is located, is also indicating that the narrow disc is disturbed by supernova interaction.

\begin{figure*}
    \includegraphics[width=1\textwidth]{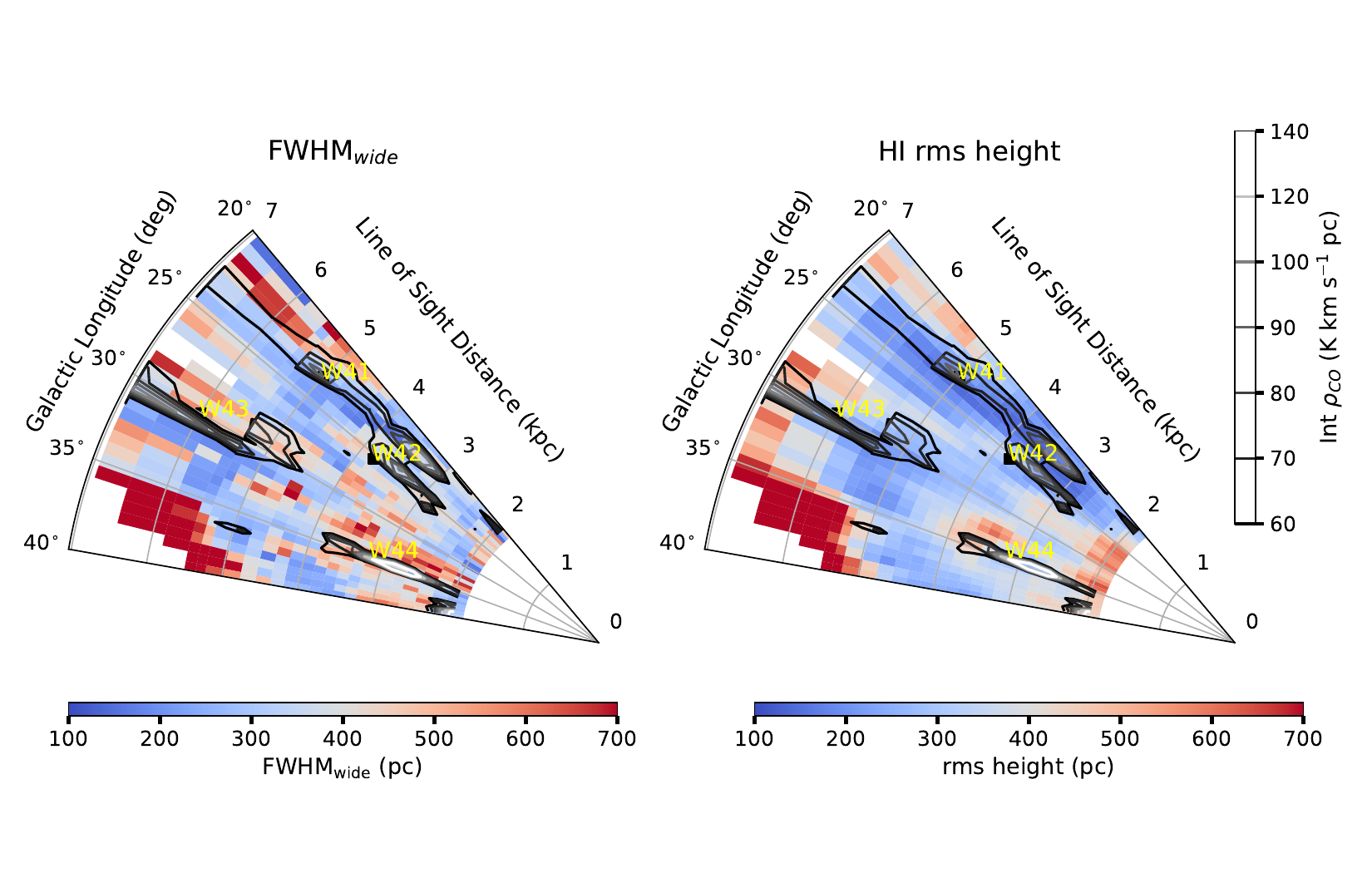}
    \caption{Density plot showing the 2D distribution of FWHM$_{\rm wide}$ (left) and rms height (right) in the entire W41-W44 region in polar coordinates where the radial axis is the line of sight distance from the Sun in units of kpc and angular coordinate is the Galactic longitude in degrees (Note that for the display purpose, the angular coordinate is scaled by a factor of two). The integrated $\rm \rho_{CO}$ levels are plotted as contours and the values of the respective contour are mentioned in the color bar. The locations of the four complexes are also annotated in the figure.}
    \label{fig:3D}
\end{figure*}

Each of these slices is separated by an average distance of $\sim$ 870 pc. Therefore, the identified troughs within a particular slice are attributed to the cumulative effect of all the sub-complex structures contained within the complex which are distributed over this 870 pc along the line of sight. A close look at the $^{12}$CO spectrum in Figure~\ref{fig:Fig2} reveals that the peak $^{12}$CO emissions from complexes  W41, W42, and W44 are falling at the edges of the velocity bins.  Hence the possible constraining effect in these complexes may not be prominently visible in these slices. 
To resolve this, a two-dimensional distribution of the \HI disc thickness must be estimated, as a function of  Galactic longitude, $l$ and line-of-sight distance, D. Subsequently, this estimated distribution of \HI disc thickness should be compared to the CO content within each respective complex.
We have attempted to estimate the two-dimensional distribution of the \HI disc thickness in the W41-W44 region using the following procedure.
We integrate the \HI and $^{12}$CO emissions along the same velocity range of 30 to 120 $\kms$ in multiple bins, now each has a smaller width of $3 \ \kms$ and hence generated numerous  slices. Further, a similar exercise was repeated to estimate the \HI disc thickness for each slice. With the known model of rotation curve as discussed in appendix~\ref{App_Dis}, we estimate the line of sight distance to each slice and approximate separation between the adjacent slices is 150 pc.
The FWHM of the \HI wide component, as a function of Galactic longitude $l$ and D, is presented as a density map in the left panel of Figure~\ref{fig:3D}. The right panel displays the density map of the disc thickness estimated from \HI rms height for the entire W41-W44 region. The radial coordinates of this polar plot represent the line-of-sight distance from the Sun in units of kpc, while the angular coordinate denotes the Galactic longitude. The superimposed contours plotted on top represent the integrated $\rm \rho_{CO}$ values at the respective positions.
Therefore, this plot is an equivalent representation of the face-on view of the W41--W44 region. The close alignment between the high-density CO contours and the locations where FWHM$_{\rm wide}$ are small around W41, W42 and W43 provides further confirmation of our findings. The prominent decrease in the \HI rms height around W41, W42 and W43  supports the anti-correlation trend that was seen between the high-density H$_2$ and \HI disc thickness previously.
Interestingly, despite the presence of high-density molecular content in the W44 region, no hint of a pinching effect was detected. This suggests that strong local stellar feedback \citep{2013ApJ...774...10S} suppresses the constraining effect in this region.

\section{Discussion and Conclusion}
\label{Sec:Con} 
We report for the first time, the observational evidence for the local vertical constraining in \HI by the Giant Molecular Cloud complexes in the W41--W44 region.
The local corrugations in the thickness of the diffuse \HI disc detected in various velocity slices are verifying the theoretical predictions by \cite{ChandaP1}. 
The narrow component represents a distribution of compact dense \HI gas along the mid-plane of the disc and no prominent correlation between its thickness and CO emission is detected. However, we see a few indications, where the FWHM of the narrow component is enhancing around W41 and W43, possibly caused by feedback from stellar activity. 
Though such local environmental activities could dominate the effect of potential  of molecular complex at smaller heights from the mid-plane, at larger heights where the diffuse \HI gas is present, gravitational attraction by the complex dominates. 
The strong negative correlation (i.e., $-0.94 < S_r < -0.6$) between the rms height and CO intensities in the entire W41--W44 region is quantitatively emphasizing this. 

Theoretically, a constant density complex shows the strongest pinching 
effect at its centre of the complex along the mid-plane\citep{ChandaP1}. 
However, for a real disc, the molecular cloud complex consists of several sub-complexes, each located 
at a different distance from the Sun and each with a different extent and density. 
Being located inside the Galaxy, from observations it is difficult to disentangle the constraining effect by individual complexes. 
Various trough-like distributions detected in each slice (as listed in table~\ref{Tab2}), are the convoluted contribution by such sub-complexes which are spread across $\sim 800 - 1000 $ pc distance along the line of sight.
The W43 GMC complex contributes to the $\12CO$ emission in velocity range, $\rm 60 < V_{LSR} < 120 \kms$, 
representing distinct molecular sub-complexes located at distances ranging from $4$ kpc to $7$ kpc from us. The trough detected
around $l\sim 30^{\circ}-32.5^{\circ}$ at different slices in these velocities gives the cleanest evidence for the constraining effect. Similarly, the GMC complex associated with W41 shows strong confirmation for the pinching effect in the same velocity ranges. 
The width of the various troughs given in table~\ref{Tab2} depends on the molecular mass distribution in the respective longitudes and the massive distribution by W41 and W42--Back creates the most wide trough.

The uniform binning along the LSR velocity ranges reduces the complexity and uncertainties that can be introduced in the transformation of Galactic coordinates to physical length scales. However, since different GMC complexes are located at slightly different  distances from the Sun, it is possible that peak CO emission of individual clouds aligns with the edges of the adjacent velocity bins. For instance, in figure~\ref{fig:Fig2}
, the W44 complex has its dominant $\12CO$ content around $45 \kms$, which is located at the margin of Slice 1 and 2. As discussed in section~\ref{Sec:Res}, unlike other GMC complexes, evidence for  pinching is not prominent for W44.   However, the two-dimensional distribution of FWHM of \HI wide component and \HI rms height estimated as in figure~\ref{fig:3D}, rules out this limitation of the velocity binning and we see that irrespective of the presence of very high density $\12CO$ content in this complex, any hint of constraining effect in W44 is not detected even in the wide component of \HI disc. This indicates that the local stellar feedback acting in this region is dominant over the gravitational potential that the complex imposes.

\begin{itemize}
\item The \HI vertical density distribution  in W41--W44 region ($20^{\circ} < l < 40^{\circ}$) which spans over around $5$ kpc along the line of sight is modelled; high-density atomic gas located around the mid-plane is found to be embedded in a  diffuse low-density gas distribution present at larger height from the Galactic plane.

\item By inspecting the multiple regions located at different distances from us, we infer the 2-D distribution of vertical thickness of  \HI disc in W41--W44. The typical variation of the thickness of the \HI wide component ranges from $200$ pc to $700$ pc, whereas that of the narrow disc is between the $50$ pc and $200$ pc. 

\item We see that the diffuse \HI gas distributed around the Galactic plane is highly sensitive to the gravitational potential imposed by the various GMC complexes and results in the trough-like features  in the FWHM of the \HI wider component. This eventually lead to local corrugations in the \HI scale height distribution. 

\item The  W43 and W41 are the dominant GMC complexes that strongly show the evidence for pinching effect in the form of a trough-like feature in scale height. The constraining effect by W43 is visible from $29^{\circ}$ to $32^{\circ}$ in Galactic longitudes and spans over a $3$ kpc distance along the line of sight of direction. The GMC complex associated with W41 shows a pinching effect in similar line-of-sight distances and in the longitude range from $21^{\circ}$ to $25^{\circ}$. 

\item The typical width of the trough varies from $\sim150$ to $\sim 850$ pc and it highly depends on the 
molecular hydrogen content in the respective velocity range. On average, the disc thickness decreases by a factor of two to three near the minima of the trough, with respect to the thickness of the \HI disc in the unaffected regions (i.e.,  $\sim700-800$ pc).  

\item The biggest trough that was detected in our analysis is created by the $6.4 \times 10^6 \Msun$ massive elongated molecular gas distribution mainly contributed by the complex W41 and a molecular cloud W42--Back. It has an approximate width of $850$ pc and a height $350$ pc. 

\item The smallest GMC that caused the pinching effect is a physically not connected molecular cloud/complex located in front of the W43. The  $2 \times 10^5 \Msun$ massive gas clouds create one of the smallest troughs detected with a width of $165$ pc and a height of $200$ pc. This enhances
the possibility of detection of effect in low mass 
molecular clouds or complexes.

\item The narrow component of the \HI disc is found to be not affected by the pinching, in fact, there are a few indications it can get disturbed by the stellar feedback in the star-forming complexes.

\item The \HI rms height which characterizes the overall extent of the disc is found to be anti-correlated (i.e., $S_r \lesssim -0.6$) with molecular gas content, in the entire W41--W44 region, for distance $3$ kpc onwards. This depicts the global constraining effect caused by the molecular gas distribution in the mid-plane.

\end{itemize}

A similar constraining effect would also be expected to be seen in stars, as
shown by \citep{ChandaP1}. Thus, such massive molecular cloud complexes
and the associated constraining of stars and \HI gas around each of these 
would constitute a substantial local perturbation to the gravitational field
in the disc. This can further affect galactic disc dynamics, for example, it
could lead to the heating of stellar velocity dispersion \citep{ChandaP1}.   Similar molecular cloud complexes are also seen in other galaxies, for example, in M83 \citep{1991Lord&Kenney}, M100 \citep{1995Rand}, and M81 \citep{1998Brouillet}. Hence the locally non-uniform or corrugated distribution of \HI and stars
predicted by \citet{ChandaP1} is a generic phenomenon. 
Our findings reveal a possibility of probing these effects in other cloud complexes
within the Galaxy, as well as in external galaxies. However, the complexity in modelling 
the 3-D distribution of gas within the Milky Way, makes the probing attempt challenging. 
Nevertheless, being the only system where such a local dynamical effect can be investigated
in detail, similar investigations need to be carried out in another part of the Galaxy.  

\section*{Acknowledgement}
  MN acknowledges postdoctoral fellowship support from a Max Planck--India Partner Group Grant and support from the Indian Institute of Science, Bangalore. NR acknowledges support from Max-Planck-Gesellschaft through
Max Planck India partner group grant. CJ would like to thank the Indian National Science Academy (INSA) for a Senior Scientist position. Authors thank the anonymous referee for the suggestions that have improved the presentation of the paper significantly.

\section*{DATA AVAILABILITY}
We use the publicly available observation data in this analysis. \HI data is from the HI4PI survey available on the website \href{https://cdsarc.u-strasbg.fr/viz-bin/qcat?J/A+A/594/A116}{https://cdsarc.u-strasbg.fr/viz-bin/qcat?J/A+A/594/A116} and $\12CO$ data in \href{https://dataverse.harvard.edu/dataverse/rtdc}{https://dataverse.harvard.edu/dataverse/rtdc}.

\bibliographystyle{mnras}
\bibliography{References}
\appendix

\section{Kinematic distance estimation}
\label{App_Dis}
Through observations, one measures the LSR velocity, $V_{LSR}$, of the gas located at the Galactic coordinates $(l,b)$ which is assumed to circularly rotate around the Galactic centre with a velocity of $V(R)$, where $R$ is the Galactocentric distance to the gas. The observed LSR velocity can be given as,
\begin{equation}
    \label{eq:VLSR}
    V_{LSR} =  \left[ \frac{R_0}{R} V(R) - V_0 \right] \sin l \cos b 
\end{equation}
     where $ R^2=D^2+R_0^2-2DR_0\cos l$.
     Here $\rm D$ is the line of sight or kinematic distance to the gas from the Sun, $\rm R_0$ and $\rm V_0$ are the Galactic constants.   Note that the above relation is purely valid in the assumption that the source is lying within the Galactic plane ($b=0^{\circ}$). In our analysis, we only consider the gas emission up to height $\lesssim 1.1$ kpc in every slice and this corresponds to a maximum angular scale of  $30^{\circ}$ (in nearer most Slice 1). Within this angular scale, we can safely assume that the distance to the gases located at different heights above the mid-plane remains the same.

With a known model for the rotation curve of the Milky Way, $V(R)$, this distance can be estimated with a unique solution in the outer Galaxy ($\rm R > R_0$) while in the inner galaxy it has two solutions $\rm D = R_0 \cos l \pm \sqrt{R^2-R_0^2\sin^2 l}$ for a particular $V_{LSR}$.
     This means that a gas component located at two discrete distances along the line of sight direction will have the same LSR velocity. This near-far ambiguity in the kinematic distance estimation is a major challenge in mapping the 3D distribution of the gas in the  Inner Galaxy. An exception is at the tangent point where $\rm D=R_0\cos l$, the kinematic distance can be estimated uniquely. 
In this analysis, we use Clemens' rotation curve model \citep{1985ApJ...295..422C}, which considers a seven-order polynomial for $V(R)$ in the inner Galaxy (we use the polynomial coefficient mentioned in Table 1 of \cite{2003PASJ...55..191N} for modelling the inner rotation curve). Note that in this analysis, since the region of W41--W44 is contained inside the inner Galactic region, we only estimate the line of sight distance to the sources within in $R < R_0$.    
\begin{figure}
    \includegraphics[width=.47\textwidth]{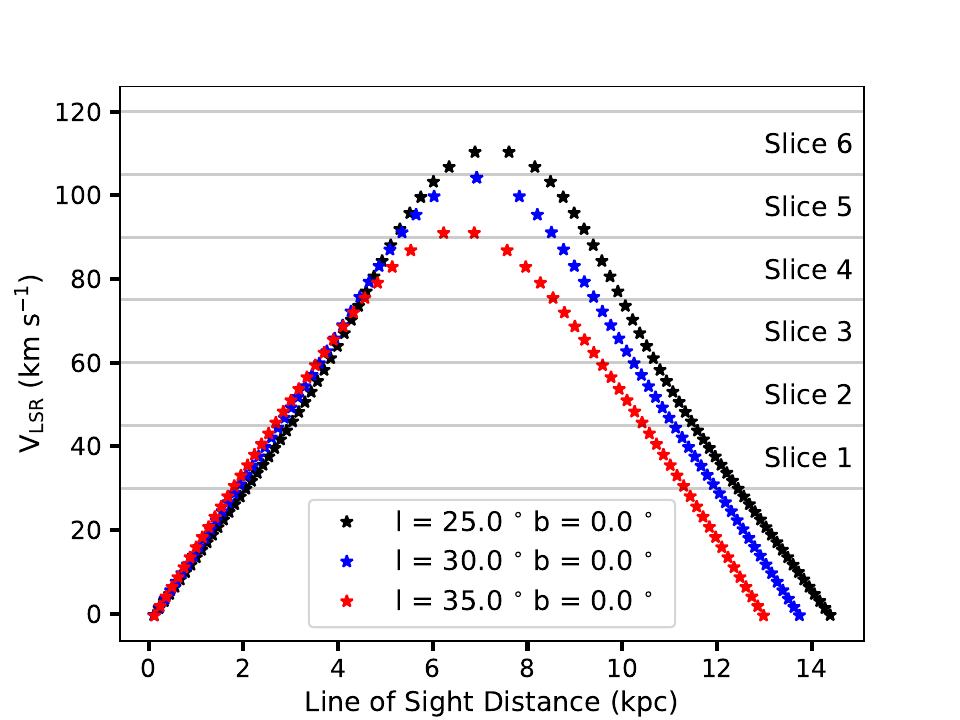}
    \caption{The LSR velocity as a function of the line of sight distance for $l= 25^{\circ},30^{\circ},35^{\circ}$. The dot-dashed horizontal lines demarcate each velocity bin. We use $R_0 = 8$ kpc and $V_0 = 220 \kms$ for the estimation.}
    \label{fig:Dist}
\end{figure}
Hence estimated kinematic distance for different Galactic longitudes is given in figure~\ref{fig:Dist}. It is clear from the figure that the near kinematic distances increase in a linear fashion with $\rm V_{LSR}$ towards the W41--W44 direction and each slice is on an average $\sim 870$ pc distance apart. Additionally, the variation of the distance along the longitude is very small in the entire region and hence we can safely assume a constant distance to the gas emission seen in the entire slice. 
\section{Disc thickness distribution for all the slices}
\label{App_SH}
The following figures show the distribution of \HI disc thickness estimated through different methods for Slice 1 -- 4 and Slice 6. 

\begin{figure*}
\includegraphics[width=.8\textwidth]{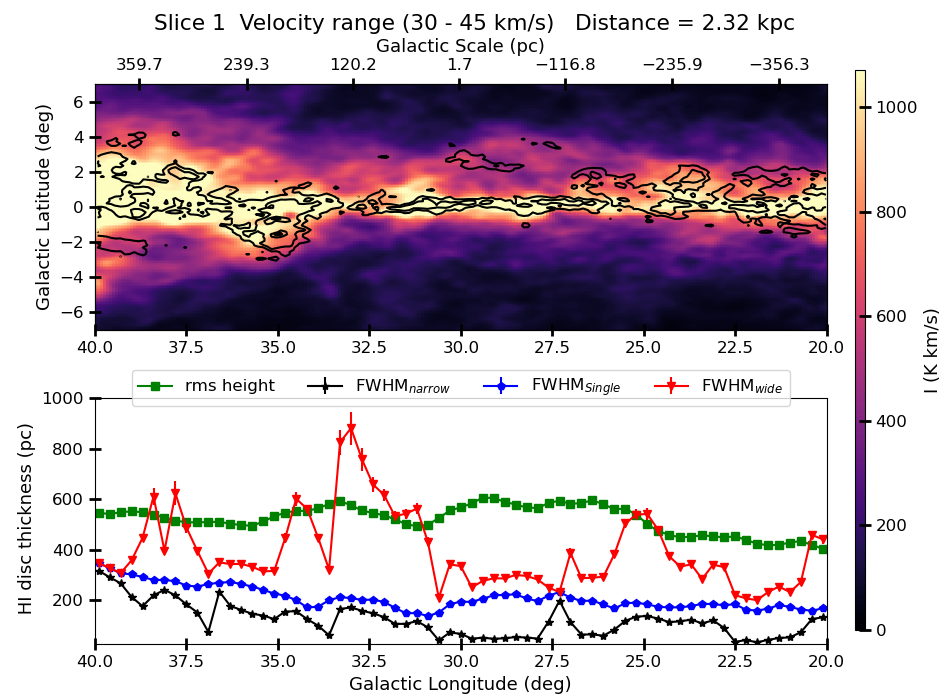}
   \caption{The distribution of \HI disc thickness in Slice 1 (Velocity range $30 - 45 \kms$). The plotting convention is similar to that of Figure~\ref{fig:Fig3}.  The values of $\12CO$ contours are $\rm 0.6, 2, 5, 7, 9 \times 8.2 \ K \kms$. The rms height plotted here is $\sigma_{h}$ multiplied by $\sqrt{8\ln 2}$, so that it act as proxy to the overall disc thickness.  }
\label{fig:SH_Slice1}
\includegraphics[width=.8\textwidth]{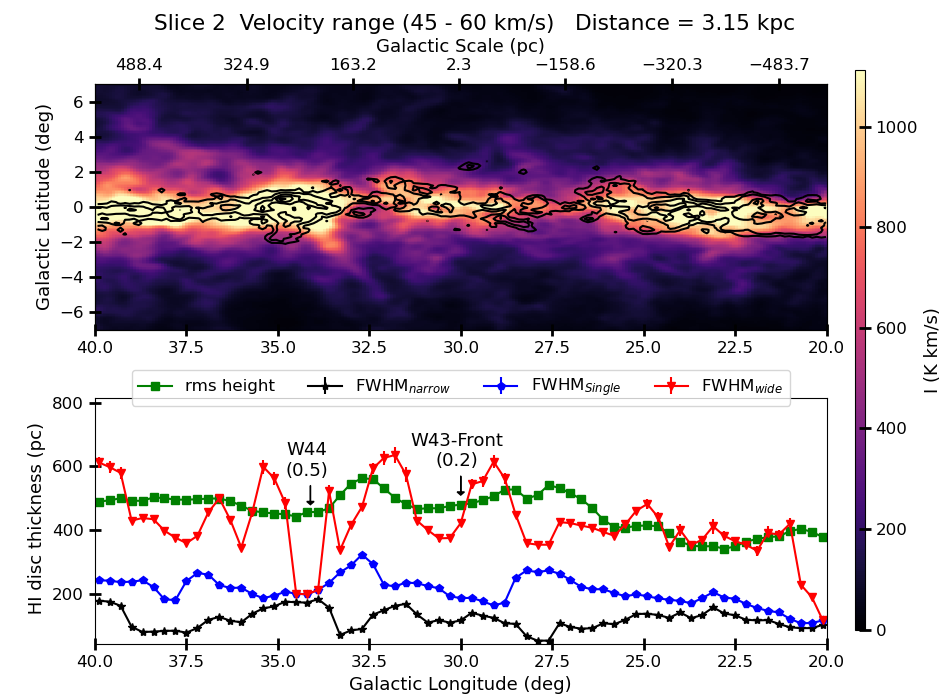}
\caption{The distribution of \HI disc thickness in Slice 2.  The values of $\12CO$ contours are $\rm 0.6, 2, 5, 7, 9 \times 6.7 \ K \kms$. The locations where the pinching effect is visible are annotated by an arrow with the name of the cloud that causing it. The total  H$_2$ mass obtained from the total $\12CO$ emission in the region of the trough is mentioned in the bracket in units of $10^6 \ \Msun$. Two troughs in the FWHM$_{\rm wide}$ is found at GMC associated with W44 and another molecular cloud located in the front of the W43, (hence `W43-Front'). }
\label{fig:SH_Slice2}
\end{figure*}

\begin{figure*}
\includegraphics[width=.8\textwidth]{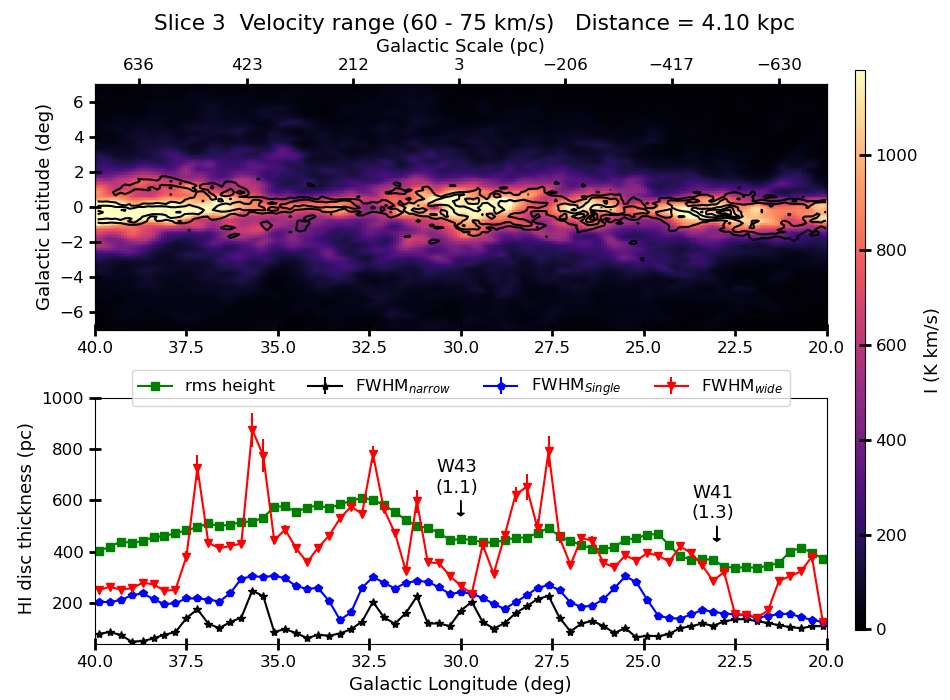}
\caption{The distribution of \HI disc thickness  in Slice 3.  The values of $\12CO$ contours are $\rm 0.6, 2, 5, 7, 9 \times 7.2 \ K \kms$. Two troughs are visible in the FWHM$_{\rm wide}$ at the location of W41 and W43.  }
\label{fig:SH_Slice3}
\includegraphics[width=.8\textwidth]{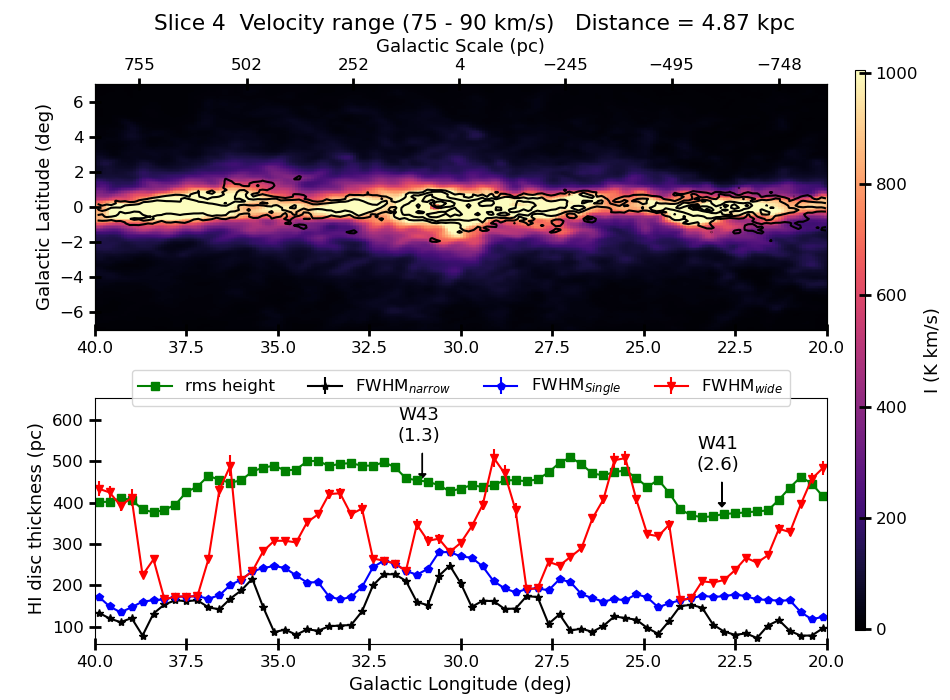}
\caption{\HI scale height distribution in Slice 4.  The values of $\12CO$ contours are $\rm 0.6, 2, 5, 7, 9 \times 8.4 \ K \kms$. Two troughs are visible in the FWHM$_{\rm wide}$, one around W41, W42--Back and the other around W43.}
\label{fig:SH_Slice4}
\end{figure*}

\begin{figure*}
\includegraphics[width=.8\textwidth]{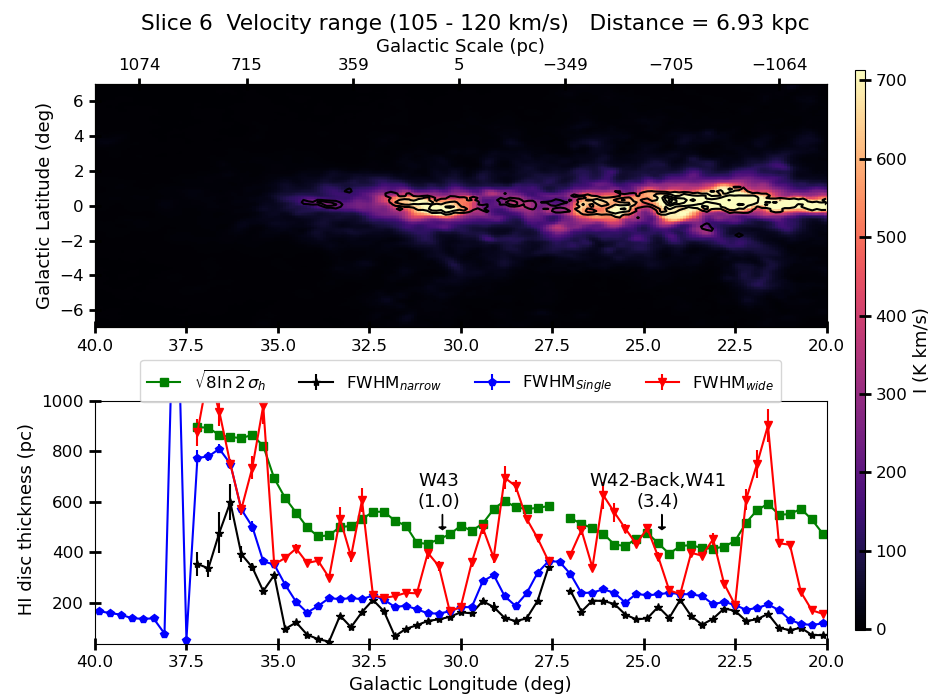}
\caption{The distribution of \HI disc thickness  in Slice 6.  The values of $\12CO$ contours are $\rm 0.6, 2, 5, 7, 9 \times 9.5 \ K \kms$. Two troughs are visible in the FWHM$_{\rm wide}$, one around W41, W42--Back and the other around W43.  }
\label{fig:SH_Slice6}
\end{figure*}

\end{document}